\documentclass{aa}
\usepackage{epsfig}
\usepackage{natbib}
\usepackage{graphicx}
\begin{document}

\newcommand{\ti}{\tilde}
\newcommand{\rw}{\rightarrow}
\newcommand{\gr}{\grave}
\newcommand{\eps}{\varepsilon}
\newcommand{\lt}{\left}
\newcommand{\rt}{\right}
\newcommand{\mc}{\mathcal} 
\newcommand{\8}{\infty}
\newcommand{\nn}{\nonumber}
\newcommand{\ra}{\vec}
\newcommand{\f}{\frac}
\newcommand{\Ll}{\ell}
\newcommand{\R}{\mathbb{R}}
\newcommand {\ltf}{\longmapsto}
\newcommand {\tf}{\mapsto}
\newcommand {\lrf}{\longrightarrow}
\newcommand{\llf}{\longleftarrow}
\newcommand {\Lrf}{\Longrightarrow}
\newcommand{\Llf}{\Longleftarrow}
\newcommand{\rf}{\rightarrow}
\newcommand{\lf}{\leftarrow}
\newcommand{\ul}{\underline}
\newcommand{\be}{\begin{equation}}      
\newcommand{\ee}{\end{equation}}      
\newcommand{\hmp}{h^{-1}Mpc}      
\newcommand{\bef}{\begin{figure}}      
\newcommand{\eef}{\end{figure}}     
\newcommand{\bea}{\begin{eqnarray}}    
\newcommand{\eea}{\end{eqnarray}}      
\newcommand{\Ga}{\Gamma}  
\newcommand{\Om}{\Offigmega}  
\newcommand{\de}{\delta}  
\newcommand{\al}{\alpha}  
\newcommand{\si}{\sigma}  
\newcommand{\bx}{{\bf x}}  
\newcommand{\lam}{\ell}  
\newcommand{\lan}{\langle}  
\newcommand{\ran}{\rangle}  
\newcommand{\etal}{{\it et al.}  }
\newcommand{\LL}{\langle \ell \rangle}  
\newcommand{\dd}{{\rm d}}
\newcommand{\kx}{\ve k\cdot \ve x} 
\newcommand{\av}[1]{\ensuremath{\left\langle #1 \right\rangle}}

\newcommand{\qmupnu}{\ensuremath{q_\mu,p_\nu}}

\def\eg{{e.g.}}
\def\ie{{i.e.}}
\def\spose#1{\hbox to 0pt{#1\hss}}
\def\ltapprox{\mathrel{\spose{\lower 3pt\hbox{$\mathchar"218$}}
\raise 2.0pt\hbox{$\mathchar"13C$}}}
\def\gtapprox{\mathrel{\spose{\lower 3pt\hbox{$\mathchar"218$}}
\raise 2.0pt\hbox{$\mathchar"13E$}}}
\def\inapprox{\mathrel{\spose{\lower 3pt\hbox{$\mathchar"218$}}
\raise 2.0pt\hbox{$\mathchar"232$}}}

\newcommand{\tdyn}{\ensuremath{\tau_\text{dyn}}}
\newcommand{\tve}[1]{\tilde{\boldsymbol{#1}}}

\def\bk{{\bf k}}  
\def\bx{{\bf x}}  
\def\bv{{\bf v}}
\def\br{{\bf r}}  
\def\ba{{\bf a}} 
\def\bde{{\mathbf\delta}} 
\def\bFo{{\bf F}} 
\def\bfo{{\bf f}} 
\def\bh{{\bf h}} 
\def\bn{{\bf n}} 
\def\bu{{\bf u}}
\def\bR{{\bf R}}
\def\mD{{\mathcal D}}
\def\nd{{\Delta\mathbf \delta}}
\def\bM{{\bf M}}
\def\bH{{\bf H}}
\def\bb{{\bf b}}
\def\bK{{\bf K}}
\def\bP{{\bf P}}
\def\by{{\bf y}}
\def\bz{{\bf z}}
\def\bS{{\bf S}}
\def\hbu{{\bf \hat\bu}}
\def\bg{{\mathbf g}}
\def\tbg{{\tilde\bg}}
\def\tmD{{\tilde\mD}}
\def\tbu{{\tilde\bu}}

\def\na{{\nabla}}
\def\nax{{\na_\bx}}
\def\naq{{\na_q}}
\def\bp{{\mathbf p}}
\def\bq{{\mathbf q}}
\def\bse{\begin{subequations}}
\def\ese{\end{subequations}}

\def\nar{{\na_\br}}
\def\da{{\dot a}}
\def\dda{{\ddot a}}
\def\bs{{\mathbf s}}
\def\bX{{\mathbf X}}
\def\bF{{\mathbf F}}

\def\bfe{{\mathbf e}}
\def\bV{{\mathbf V}}
\def\bU{{\mathbf U}}
\def\hrho{{\hat\rho}}
\def\inte{\mathrm{int}}
\def\bm{\mathbf{m}}
\def\bq{\mathbf{q}}
\def\bw{\mathbf{w}}
\def\mV{\mathcal{V}}
\def\mL{\mathcal{L}}
\def\mP{\mathcal{P}}
\def\mF{\mathcal{F}}
\def\ov{\lan v\ran}
\def\og{\lan g\ran}
\def\oF{\lan F\ran}
\def\bl{\mathbf{l}}
\def\bze{\mathbf{0}}
\def\bgp{\mathbf{g}_{pec}}
\def\lsim{\raise 0.4ex\hbox{$<$}\kern -0.8em\lower 0.62ex\hbox{$\sim$}} 
\def\gsim{\raise 0.4ex\hbox{$>$}\kern -0.7em\lower 0.62ex\hbox{$\sim$}} 
\def\mP{\mathcal{P}}
\def\mQ{\mathcal{Q}}
\def\mN{\mathcal N}
\def\mK{\mathcal K}
\def\mR{\mathcal R}
\def\bi{\mathbf{i}}
\def\bj{\mathbf{j}}
\def\bk{\mathbf{k}}
\def\tu{\tilde{u}}
\def\mH{\mathcal{H}}
\def\rhoo{{\rho^{(1)}}}
\def\rhot{{\rho^{(2)}}}
\def\ha{\hat a}
\def\dha{{\dot{\hat a}}}
\def\ddha{{\ddot{\hat a}}}
\def\dA{\dot A}
\def\ddA{\ddot A}
\def\mA{\mathcal A}
\def\fN{f^{(N)}}
\def\f0N{f_0^{(N)}}
\def\bec{\begin{center}}
\def\eec{\end{center}}
\def\tde{\tilde\de}
\def\pec{{\mathbf pec}}

\title{A toy model to test the accuracy of cosmological N-body
  simulations}

\subtitle{}

\author{Francesco Sylos Labini \inst{1,2}}

\titlerunning{Testing the accuracy of cosmological N-body simulations} 

\authorrunning{Sylos Labini}

\institute{ 
Centro Studi e Ricerche Enrico Fermi, Via Panisperna 89 A, 
Compendio del Viminale, 00184 Rome, Italy
\and 
Istituto dei Sistemi Complessi CNR, 
Via dei Taurini 19, 00185 Rome, Italy.
}

\date{Received / Accepted}

\abstract{  The evolution of an isolated over-density represents a
  useful toy model to test the accuracy of a cosmological N-body code
  in the non linear regime as it is approximately equivalent to that
  of a truly isolated cloud of particles, with same density profile
  and velocity distribution, in a non expanding background. This is
  the case as long as the system size is smaller than the simulation
  box side, so that its interaction with the infinite copies can be
  neglected. In such a situation, the over-density rapidly undergoes
  to a global collapse forming a quasi stationary state in virial
  equilibrium.  However, by evolving the system with a cosmological
  code ({\tt GADGET}) for a sufficiently long time, a clear deviation
  from such quasi-equilibrium configuration is observed.  This occurs
  in a time $t_{LI}$ that depends on the values of the simulation
  numerical parameters such as the softening length and the
  time-stepping accuracy, i.e. it is a numerical artifact related to
  the limited spatial and temporal resolutions.  The analysis of the
  Layzer-Irvine cosmic energy equation, confirms that this deviation
  corresponds to an unphysical dynamical regime.  By varying the
  numerical parameters of the simulation and the physical parameters
  of the system we show that the unphysical behaviour originates from
  badly integrated close scatterings of high velocity particles.We
  find that, while the structure may remain virialized in the
  unphysical regime, its density and velocity profiles are modified
  with respect to the quasi-equilibrium configurations, converging
  however to well defined shapes, the former characterised by a
  Navarro Frenk White-type behaviour.
 }

\maketitle

\section{Introduction} 

In order to control the numerical integration accuracy of a finite
self-gravitating system evolution in a static background one may use
total energy and angular momentum conservation \citep{aarseth}.  { 
  In the cosmological case, instead of being conserved, the total
  (peculiar) energy decreases with time.  Indeed, the cosmic energy
  equation, known as the Layzer-Irvine (LI) equation, describes the
  change of the total peculiar energy of a system due to the adiabatic
  expansion of the background \citep{peebles}.  The quantification of
  the deviations from the LI equation is one of the very first tests
  made in all cosmological codes (see, e.g.,
  \citet{teyssier,miniati}).  However, when typical initial conditions
  of standard models of structure formation are considered, the
  application of the LI test is limited by several reasons.  First of
  all, high redshift initial conditions of typical cosmological N-body
  simulations are very uniform and cold, so that both the initial
  peculiar potential and kinetic energies are close to zero: this
  makes difficult the numerical estimation of the cosmic energy
  equation. In addition, at low redshifts, the density field is
  typically characterised by a great variety of non linear structures
  of different sizes and by low density regions still in the linear
  regime.  Structures of different sizes and in different physical
  regimes give very different contributions to total the peculiar
  potential and kinetic energies so that it is not straightforward to
  estimate which error in the cosmic energy equation maybe tolerated
  and/or which constraints on the accuracy of the simulation maybe
  placed (see also discussion in \citet{jsl13} and references
  therein).  }

For this reason several authors have developed convergence tests to
explore the role of the various physical and numerical parameters that
characterise a given simulation with the aim of establishing the
reliability of the results (see e.g., \citet{power_2002}).
Alternatively there are studies that compare different algorithms (see
e.g., \citet{knebe,iannuzzi} and references therein). However, it is
not a simple issue to control the numerical accuracy of the non-linear
gravitational clustering in a { cosmological simulation}; for instance
the question of determining the optimal spatial resolution, i.e. the
best value for the gravitational force softening length, is still an
open problem (see e.g.,
\citet{splinter_1988,power_2002,romeo08,joyce_2008}).

In this paper, following the work of \citet{jsl13}, we consider the
evolution of a very simple system, namely an isolated over-density in
an expanding background, as a toy model to control the accuracy of the
numerical integration and to test the correctness of a cosmological
simulations.

The evolution of this kind of structure in a static background, has
been studied since the earliest numerical N-body experiments
\citep{henon_1964,vanalbada_1982,aarseth_etal_1988,david+theuns_1989,theuns+david_1990,bertin2000,heggie2003,boily+athanassoula_2006,jmsl09a,sl12a}.
Briefly, driven by its own mean field, it collapses in a time scale
$\tau_c \approx \sqrt{G\rho}^{-1}$, forming a virialized
quasi-stationary state (QSS), i.e. a collision-less configuration that
is not in thermodynamical equilibrium but that has a {finite} lifetime
fixed by the collisional relaxation time-scale
\be
\label{tau2}
\tau_2 \approx \kappa N \tau_c \;,
\ee 
where $N$ is the system particles
number and $\kappa$ is a numerical factor that takes into account the
``Coulomb logarithm"
\citep{chandra43,theis+spurzem_1999,diemandetal_2body,gabriellietal_prl2010}.

\citet{jsl13} have shown that an isolated over-density in { a
  cosmological simulation} should in principle evolve, in physical
coordinates, just as if it were in a non-expanding space { with open
  boundary conditions}.  This corresponds to the fact that such a
structure is in the stable clustering regime. {Namely,} it evolves as
it were { truly} isolated from the rest of universe.  While the
{stable clustering} regime is assumed to approximately describe the
evolution of a structure in the strongly non-linear regime
\citep{peebles}, in the case of a single over-density {with periodic
  boundary conditions} and negligible finite size effects it must be
precisely {  followed:} the stable clustering regime {  thus}
corresponds to a QSS for a structure in a static background.

Indeed, in a cosmological simulation, where periodic boundary
conditions are used, a single over-density can be considered isolated
if finite size effects are negligible.  {  Such} mass distribution
is isolated but for the interaction with its ``copies" included in the
infinite system over which the force is summed. It is easy to show
that the copies contribute to the force with terms that represent
perturbations suppressed by positive powers of $R_0/L$, where $R_0$ is
the over-density size and $L$ is the side of the periodic box
\citep{gabrielli_2006,jsl13}.  {  When the latter condition is
  satisfied, the only difference in the evolution between the
  expanding and the static case can arise from the smoothing of the
  gravitational force.}

{  Beyond the comparison of the expanding case with the template
  given by the static background evolution, one may consider the the
  analysis of the LI equation.  In fact, \citet{jsl13} found that is a
  useful diagnostic to control the accuracy of cosmological
  simulations when the evolution of a single over-density is
  considered.  In this case,} the breakdown of the LI test occurs when
the structure leaves the stable clustering regime: this is due to the
difficulty of numerically integrating the system evolution with
sufficient accuracy because hard two-body collisions take place.

We consider in this paper {  the evolution of } an initially
spherical {  and uniform} cloud of particles with a velocity
dispersion such that the initial virial ratio is in the range $b_0 \in
[-1,0]$.  When the system is warm enough, i.e. $b_0 \approx -1$, it
approximately maintains its original size {and shape} during and after
the collapse. Instead, when the system is cold enough, i.e. $b_0
\approx 0$, it undergoes to a stronger contraction so that its size is
considerably reduced during the collapse forming thereafter a
virialized structure with a density profile {  decaying as $r^{-4}$}
\citep{jmsl09a,sl12a}.  {  Our goal is to study the collapse of such
  a single cloud of particles in an expanding universe: by monitoring
  the LI equation and the stability of the clustering (in real space)
  we may understand to which accuracy the cosmic energy equation must
  be satisfied, in the strongly non linear regime, so that the
  simulation still provides with physically meaningful results.  In
  particular we will study the accuracy of the numerical integration
  by changing the numerical and physical parameters that characterise
  a simulation.}

The paper is organised as follows.  In Sect.\ref{evolution} we recall
the main results of the evolution of an isolated over-density in a
static background. Then we summarise the results of \citet{jsl13}
concerning the {  approximate} equivalence of an isolated cloud of
particles in a static background {  with open boundary conditions
  and the same cloud when it is placed alone in a box of an infinite
  periodic system with space expansion --- i.e., when the simulation
  is run in cosmological conditions.}  In Sect.\ref{simulations} we
discuss the preparation of the initial conditions, the choice of the
parameters that control the numerical integration and we briefly
recall the LI test.  The comparison between the evolution of an
isolated over-density in a static and expanding background is
presented in Sec.\ref{comparison_short_long}. A series of systematic
tests to determine the role of the integration parameters such as the
gravitational force softening length and the accuracy of the
time-stepping is presented in Sect.\ref{numerical_effects}. Then in
Sect.\ref{scaling_and_finite_size} we discuss several tests performed
by changing the physical parameters of the initial conditions, such as
the number of particles, the system size and the velocity
dispersion. Finally in Sect.\ref{conclusion} we discuss the principal
results and then we draw our main conclusions.




\section{Isolated over-density in a cosmological and a static background} 
\label{evolution}

Firstly we briefly recall the main features of the collapse of an
isolated over-density in a static background with open boundary
conditions (hereafter STAT case).  Then we discuss the equivalence of
the evolution of an isolated structure in STAT case and in an
expanding background with periodic boundary conditions (EXP case).

\subsection{Isolated over-densities in a static background} 
\label{static_background_case} 

In the STAT case, the global collapse of an isolated and spherical
cloud of self-gravitating particles, of density $\rho$, is {  determined} by
its mean gravitational field and it is characterised by the time scale
\be
\label{tauC}
\tau_c = \sqrt{\frac{3\pi}{32G \rho}} \;. 
\ee
When the initial velocity dispersion is such that the virial ratio is
$b_0 = 2K_0/W_0\approx -1$ (where $K_0/W_0$ is the system initial
kinetic/potential energy) the collapse consists in a few oscillations
that drive the system to a virialized QSS: {  this maintains
  approximately the size $R_0$ and shape of the initial configuration
\citep{sl12a}.}

On the other hand, when the initial velocity dispersion is small
enough (or zero), the system undergoes to a stronger contraction
reducing its size by a large factor. In this case the features of the
virialized QSS depend explicitly on the number of system particles (at
constant density): for instance, when particles are initially Poisson
distributed, the intrinsic characteristic size in Eq.\ref{nrb0} scales
as $r_c \approx N^{-1/3}$ \citep{aarseth_etal_1988,jmsl09a}.  Other
noticeable features are: (i) during the collapse a fraction of
particles gain enough kinetic energy to escape from the system; (ii)
the density profile after the collapse is characterised by the power
law decay {  described by 
\be
\label{nrb0} 
n(r) = \frac{n_c}{1+(r/r_c)^4} \;, 
\ee 
(that we  denominate the quasi equilibrium profile --- QEP) where $r_c$ and
$n_c$ are two constants. }

The key to understand the formation of this density profile is
represented by the physics of the ejection mechanism.  The main
characteristics of the virialized QSS that can be framed in {  a
  simple physical model introduced by \citet{sl12a} } are
\bea
\label{mainb1}
 && 
n(r) \sim r^{-4} \; , \\ &&
\label{mainb2}
\sigma^2_r(r) \equiv \langle v_r^2 \rangle \sim r^{-1} \;, \\ &&
\label{mainb3}
\Phi(r) \equiv \frac{n(r)}{\sigma^3_r(r)} \sim \frac{r^{3/2}}{r^4}
\sim r^{-5/2} \;, \eea
where $v_r$ is the radial velocity and $\Phi(r)$ is {  the so-called
  pseudo phase-space density}.  Note that, once the QSS is formed, its
potential and kinetic energy reach a {\it constant value}: this fact
represents one of the diagnostics to numerically detect the quasi
stationarity of the particle configuration
\citep{jmsl09a,sl12a,jsl13}.


\subsection{Equivalence of the evolution in a static and in an 
expanding background} 

\label{expanding_background_case}

\citet{jsl13} have {  shown} that when a cosmological code is used
to simulate an isolated structure (of size $R_0$) in an expanding
universe, it should reproduce the same results, in physical
coordinates, as that obtained for the same structure in open boundary
conditions without expansion. Let us briefly recall in which limit
this equivalence is verified.

Dissipation-less cosmological N-body simulations solve numerically the
equations \citep{peebles}
\begin{equation}
\frac{d^2 {\bf x}_i}{dt^2} + 2H \frac{d{\bf x}_i}{dt} = -
\frac{1}{a^3} Gm \sum^{copies} \sum_{j \neq i}^{j=1,..N} \frac{{\bf x}_i -
  {\bf x}_j}{\vert {\bf x}_i - {\bf x}_j \vert^3}
\label{3d-equations-1}
\end{equation}
where ${\bf x}_i$ are the comoving particle coordinates of the
$i=1,...,N$ particles of equal mass $m$, enclosed in a cubic box of
side $L$, and subject to {\it periodic boundary conditions}, $a(t)$ is
the appropriate scale factor for the cosmology considered, and
$H(t)={\dot a}/{a}$ is the Hubble constant. Note that the force on a
particle is that due to the $N-1$ other particles {\it and to all
  their copies}.  {  The leading non-zero term of the force due to
the copies is a repulsive ``dipole''; by neglecting the quadrupole and
higher multipole terms, that are convergent and suppressed by positive
powers of $(R_0/L)$ compared to the dipole term
\citep{gabrielli_2006,jsl13}, we can rewrite Eq.\ref{3d-equations-1}
as} 
\begin{equation}
\frac{d^2 {\bf x}_i}{dt^2} + 2H \frac{d{\bf x}_i}{dt} =
-\frac{Gm}{a^3} \sum_{j \neq i}^{j=1,..N} \frac{{\bf x}_i - {\bf x}_j}{\vert
  {\bf x}_i - {\bf x}_j \vert^3} + \frac{4\pi G \rho_0}{3} \bx_i
\label{3d-equations-3}
\end{equation}
where the sum is now only over the $N-1$ particles in the simulation box.
Assuming for simplicity an Einstein-de Sitter (EdS) cosmology, for which 
\begin{equation}
\frac{\ddot{a}}{a} = - \frac{4\pi G \rho_0}{3 a^3}\,,
\end{equation}
Eq.\ref{3d-equations-1} may be written, in physical coordinates $\br_i
\equiv a(t) \bx_i$, simply as
\begin{equation}
\frac{d^2 {\bf r}_i}{dt^2} = -Gm \sum_{j \neq i}^{j=1,..N} \frac{{\bf r}_i -
  {\bf r}_j}{\vert {\bf r}_i - {\bf r}_j \vert^3} \;.
\label{3d-equations-4}
\end{equation}
These are the equations of motion of $N$ purely
self-gravitating particles.  Therefore, up to finite size corrections
that vanish in the limit $R_0/L \rightarrow 0$, the equations of
motion (in physical coordinates) in an expanding universe are the same
of those describing the evolution {  of the same finite structure}
in a static background with open
boundary conditions.

We stress that Eq.\ref{3d-equations-4} represents an important
  approximation that justifies the analogy between the same cloud of
  self-gravitating particles in STAT and EXP conditions.  This is not
  a trivial approximation as it consists in replacing the sum in
  Eq.\ref{3d-equations-1} on a infinite periodic system, i.e. with
  infinite terms, with the sum on a finite system in
  Eq.\ref{3d-equations-4}. The periodic sum can be written as a local
  sum on $N$ particles and a non-local term which takes into account
  the contribution of the infinite copies: this latter term can be
  neglected for the system considered, i.e. for $R_0<L$.  Only in this
  situation the evolution of a single over-density in a periodic
  system, i.e., an infinite number of over-densities one in each of
  the infinite boxes of the periodic system, 
  {  in an expanding background}  is
  equivalent to the evolution of a truly isolated over-density in a
  static background with open boundary conditions.

Note that the simulations are run, as usual, in comoving
coordinates. In what follows we present analyses both in comoving and
in physical coordinates because the stable clustering regime
should manifest itself in physical coordinates, i.e.  the structure
remains almost invariant only in these coordinates \citep{jsl13}.




\section{The simulations}

\label{simulations}

\subsection{Initial conditions}
\label{initial_conditions}

We have considered a system with total mass $M$ in a spherical volume
$V=4 \pi/3 R_0^3$: this can be discretized by using $N$ randomly
distributed particles of mass $m$ such that 
\be
\label{density_eq}
\rho = \frac{M}{V} = \frac{ 3 N m}{4 \pi R_0^3}= \mbox{const.}  
\ee 
We have used $10^3 \le N<10^4$ (series S), $N=10^4$ (series M) and $10^4
<N \le 10^5 $ (series L).  Details of the simulations are reported in
Table~\ref{table}: all length scales are given in units of the box
side $L$. Note that the results we report require only choice of units
for length and energy: for the former we will take units defined by
$L=1$, and for the latter units in which $W_0 = 3GM^2/(5R_0)=-1$,
i.e. in which the initial continuum limit gravitational potential
energy is minus unity.

\begin{table}
\begin{center}
\begin{tabular}{|c|c|c|c|c|c|c|}
\hline Name  &  $N$     &$|b_0|$&   $R_0$  &  $\varepsilon$         & $\varepsilon/\ell$  &$\eta$  \\
\hline 
     S1a     & 1E3  &0     &  1.0E-1     &3.7  E-6 & 4.2E-4    &    2.5E-2    \\ 
\hline
     S4a     & 1E3  &0     &  1.0E-1     &3.7  E-5 & 4.2E-3    &    2.5E-2    \\ 
     S4b     & 2E3  &0     &  1.0E-1     &3.7  E-5 & 5.3E-3    &    2.5E-2    \\ 
     S4c     & 4E3  &0     &  1.0E-1     &3.7  E-5 & 6.6E-3    &    2.5E-2    \\ 
     S4d     & 6E3  &0     &  1.0E-1     &3.7  E-5 & 7.6E-3    &    2.5E-2    \\ 
     S4e     & 8E3  &0     &  1.0E-1     &3.7  E-5 & 8.3E-3    &    2.5E-2    \\ 
     S4f     & 1E3  &0     &  1.0E-1     &3.7  E-5 & 8.3E-3    &    2.5E-3    \\ 
     S4g     & 1E3  &0     &  1.0E-1     &3.7  E-5 & 8.3E-3    &    5.0E-4    \\ 
\hline
     S6a     & 1E3  &0     &  1.0E-1     &3.7  E-4 & 4.2E-2    &    2.5E-2    \\  
\hline
\hline
  M1a        & 1E4  &0     &  1.0E-1     &3.7  E-6            & 9.0E-4   &    2.5E-2    \\ 
  M1b        & 1E4  &0     &  1.0E-1     &3.7  E-6            & 9.0E-4   &    5.0E-3    \\ 
  M1c        & 1E4  &0     &  1.0E-1     &3.7  E-6            & 9.0E-4   &    1.3E-2    \\ 
  M1d        & 1E4  &0     &  1.0E-1     &3.7  E-6            & 9.0E-4   &    2.5E-2    \\
  M1e        & 1E4  &0     &  1.0E-1     &3.7  E-6            & 9.0E-4   &    1.3E-1    \\
  M1f        & 1E4  &0     &  1.0E-1     &3.7  E-6            & 9.0E-4   &    2.5E-1    \\
\hline
  M2a        & 1E4  &0     &  1.0E-1     &7.4  E-6            & 1.8  E-3   &    2.5E-3   \\  
\hline
  M3a        & 1E4  &0     &  1.0E-1    &1.0  E-5             & 2.4  E-3  &    2.5E-2    \\  
\hline
 M4a      & 1E4  &0     &  1.0E-1     &3.7  E-5               & 9.0   E-3   &    2.5E-2     \\ 
 M4b      & 1E4  &0     &  1.0E-1     &3.7  E-5               & 9.0   E-3   &    2.5E-2     \\ 
 M4c      & 1E4  &0     &  1.0E-1     &3.7  E-5               & 9.0   E-3   &    2.5E-2     \\ 
 M4d      & 1E4  &0.25  &  1.0E-1     &3.7  E-5               & 9.0   E-3   &    2.5E-2     \\ 
 M4e      & 1E4  &0.5   &  1.0E-1     &3.7  E-5               & 9.0   E-3   &    2.5E-2    \\  
 M4f      & 1E4  &0.75  &  1.0E-1     &3.7  E-5               & 9.0   E-3   &    2.5E-2     \\ 
 M4g      & 1E4  &1.0   &  1.0E-1     &3.7  E-5               & 9.0   E-3   &    2.5E-2    \\  
 M4h      & 1E4  &0     &  2.5E-2     &3.7  E-5               & 3.7   E-2   &    2.5E-2     \\ 
 M4k      & 1E4  &0     &  5.0E-2     &3.7  E-5               & 1.7   E-2   &    2.5E-2     \\ 
 M4i      & 1E4  &0     &  2.0E-1     &3.7  E-5               & 4.4   E-3   &    2.5E-2    \\  
 M4j      & 1E4  &0     &  3.0E-1     &3.7  E-5               & 3.0   E-3   &    2.5E-2     \\ 
 M4l      & 1E4  &0     &  1.0E-1     &3.7  E-5               & 9.0   E-3   &    2.5E-3     \\ 
 M4m      & 1E4  &0     &  1.0E-1     &3.7  E-5               & 9.0   E-3   &    5.0E-3     \\ 
 M4n      & 1E4  &0     &  1.0E-1     &3.7  E-5               & 9.0   E-3   &    5.0E-2     \\ 
 M4o      & 1E4  &0     &  1.0E-1     &3.7  E-5               & 9.0   E-3   &    1.0E-1     \\ 
\hline
M5a       & 1E4  &0     &  1.0E-1     &7.4  E-5               & 1.8   E-2   &    2.5E-2     \\ 
\hline
M6a          & 1E4  &0     &  1.0E-1  &3.7  E-4               & 9.0    E-2  &    2.5E-2      \\ 
M6b          & 1E4  &0     &  2.5E-2  &3.7  E-4               & 3.7    E-1  &    2.5E-2      \\ 
M6c          & 1E4  &0     &  5.0E-2  &3.7  E-4               & 1.8    E-1  &    2.5E-2      \\ 
M6d          & 1E4  &0     &  2.0E-1  &3.7  E-4               & 4.5    E-2  &    2.5E-2      \\ 
M6e          & 1E4  &0     &  3.0E-1  &3.7  E-4               & 3.0    E-2  &    2.5E-2      \\ 
\hline%
M7a         & 1E4  &0     &  1.0E-1   &7.4  E-4               &  1.8   E-1  &    2.5E-2     \\ 
\hline
M8a         & 1E4  &0     &  1.0E-1   &3.7  E-3               & 9.0    E-1  &    2.5E-2    \\  
\hline
\hline
L4a         & 2E4  &0     &  1.0E-1     &3.7  E-5             & 1.1    E-2   &    2.5E-2    \\  
L4b         & 4E4  &0     &  1.0E-1     &3.7  E-5             & 1.4    E-2   &    2.5E-2    \\  
L4c         & 6E4  &0     &  1.0E-1     &3.7  E-5             & 1.6    E-2   &    2.5E-2    \\  
L4d         & 8E4  &0     &  1.0E-1     &3.7  E-5             & 1.8    E-2   &    2.5E-2    \\  
L4e         & 1E5  &0     &  1.0E-1     &3.7  E-5             & 1.9    E-2   &    2.5E-2    \\  
\hline 
\end{tabular}
\end{center}
\caption{Details of the N-body simulations: $N$ is the number of
  particles in the spherical over-density, $|b_0|$ is the absolute
  value of its initial virial ratio, $R_0$ is its initial radius,
  $\varepsilon$ is the softening length, $\ell$ is the initial average
  distance between nearest neighbours and $\eta$ is the parameter that
  control the accuracy of the time-stepping.  Only in the case of the
  simulation M4b (M4c) we have changed the parameter $\theta$ from 0.7
  to 0.1 (the parameter $\phi$ from $5 \times 10^{-3}$ to $5 \times
  10^{-4}$). (See Sect.\ref{numerical_parameters} for a discussion of
  the various numerical parameters of the code, i.e. $\varepsilon,
  \eta, \theta, \phi$).}
\label{table}
\end{table}
 Note that the system radius $R_0$ is chosen to be $R_0 < L/4$ so to
 minimise, as discussed above, the effect of the periodic boundary
 conditions. {In particular, the distance between any pair of
   structure particle $i,j$ is less than that separating $i$ from any
   particle of the copies in the infinite periodic system.}

The spherical cloud corresponds, in a cosmological simulation, to an
over-density with initial mass density contrast
\begin{equation}
\delta = \frac{\rho}{\rho_0} = \frac{3L^3}{4\pi R_0^3} \;,
\label{ratio-densities}
\end{equation}
where $\rho=3M/(4\pi R_0^3)$ is the cloud density and $\rho_0 = M/L^3$
is the mean mass density of the universe (see Tab.\ref{tableErrE}).  
Particles mass is such that $\rho_0$ is equal to the critical density. 
\begin{table}
\begin{center}
\begin{tabular}{|c|c|c|c|}
\hline 
$R_0$  &  $\delta$ & $4/(3 \pi)\sqrt{\delta}$ & $\ell$ \\
\hline 
0.025 &  15360     & 52.6                     & 1.0E-3   \\
0.05  &  1920      & 18.6                     & 2.1E-3   \\
0.1   &  240       & 6.6                      & 4.1E-3   \\
0.2   &  30        & 2.3                      & 8.2E-3   \\
0.3   &  8.9       & 1.1                      & 1.2E-2   \\
\hline 
\end{tabular}
\end{center}
\caption{Parameters of the spherical clouds with different initial
  radius $R_0$ and with over-density $\delta$
  (Eq.\ref{ratio-densities}). The third column reports the pre-factor
  of Eq.\ref{time-elapsed}.  The average distance between nearest
  neighbours $\ell$ is computed for $N=10^4$, i.e. $\ell \approx 4.12\times
  10^{-2} R_0$.}
\label{tableErrE}
\end{table}

The scale factor in an EdS universe obeys to \citep{peebles}   
\be
\label{sf_eds}
a(t) = a_0 \left( \frac{t}{t_0} \right)^{2/3} \;, 
\ee 
where
$t_0=1/\sqrt{6 \pi G \rho_0}$. Assuming $a_0=1$ from
Eqs.\ref{ratio-densities}-\ref{sf_eds} we find \citep{jsl13}
\begin{equation}
t-t_0=[a^{3/2}-1] t_0 = \frac{4}{3\pi} \sqrt{\delta} \, [a^{3/2} -1]
\tau_{c}
\label{time-elapsed}
\end{equation}
where $\tau_c$ is given by Eq.\ref{tauC}.  {  Note that from
  Eq\ref{time-elapsed} we find} that only for very high values of the
over-density, i.e. $\delta >10^4$, at moderate expansion factors $a
\approx 10$, $(t-t_0)/\tau_c$ is of the order of a thousand i.e. of
the order of $\tau_2$ for $N=10^3$ (see Eq.\ref{tauC}).




\subsection{Numerical integration parameters} 
\label{numerical_parameters} 

We have used the publicly available tree-code {\tt GADGET}
\citep{gadget_paper,springel_2005}. In this code one may choose the
values of a number of parameters and we have followed the
prescriptions of the user guide for {\tt GADGET-2 }\footnote{see {\tt
    http://www.mpa-garching.mpg.de/gadget/users-guide.pdf}} to fix the
basic ones.

The gravitational potential used by {\tt GADGET} has the exact
Newtonian potential above a separation of $2.8 \varepsilon$ while for
smaller separations the potential first derivative vanishes
\footnote{the value of $\varepsilon$ is the value of the parameter
  with this name in the code}.  The different values of $\varepsilon$
used are reported in Tab.\ref{table}. {  
Note that we have chosen a softening length that 
is fixed in comoving coordinates and thus it varies in physical ones.}

Further we have set, following \citet{power_2002,springel_2005} a
Barnes-Hut opening criterion with an opening angle of the tree
algorithm $\theta=0.7$ for the first force computation and a dynamical
updating criterion thereafter~\footnote{but $\theta=0.1$ in the
  simulation M4b}. The accuracy of the relative cell-opening criterion
(i.e., the parameter {\tt ErrTolForceAcc})~\footnote{but $\phi = 5
  \times 10^{-4}$ in the simulation M4c} is set to $\phi = 5 \times
10^{-3}$.  Finally, we have chosen the time-stepping criterion 0 of
{\tt GADGET} where the value of the parameter controlling its accuracy
$\eta$ (i.e., the parameter {\tt ErrTolIntAccuracy}) is reported in
Tab.\ref{table}.


\subsection{Basic definitions}

Let us introduce some basic definitions that will be useful in what follows. 
The kinetic
energy in physical coordinates is
\be
\label{Kp}
K_p = \frac{1}{2} \sum_{i=1,N} m \left(\frac{d {\bf r}_i} {dt} \right)^2
\ee
and the potential energy in physical coordinates is 
\be
\label{Wp}
W_p =  \frac{1}{2} \sum_{i=1,N} m_i  \Phi({\bf r}_i)
\ee
where 
\be 
\Phi({\bf r}_i) = - G \sum_{j=1,N}^{j \ne i} m_j g(| {\bf r}_i - {\bf r}_j|) 
\ee
and  where $g(r)$ is the exact {\tt GADGET} pair potential 
\citep{springel_2005}.  
The virial ratio is defined as 
\be
\label{b}
b = \frac{2 K_p}{W_p} \;. 
\ee
Note that the initial virial ratio is $b_0=2K_0/W_0$, where $K_0, W_0$ are 
respectively the initial kinetic and potential energy. 
We define the potential energy in comoving coordinates  
\be
\label{Wc} 
W_c=\frac{1}{2} \sum_{j} m_j \Phi({\bf x}_i)  = W_p \times a 
\ee 
and the {\tt GADGET} gravitational potential energy 
\be
\label{Wg} 
W_g=\frac{1}{2} \sum_{j=1,N}^{copies} m_j \Phi({\bf x}_i)  \;,
\ee
i.e., the two-body potential is calculated by using periodic boundary
conditions. As mentioned above, in the limit 
$R_0 \ll L$, we have $W_g\approx W_c$. 

The peculiar kinetic energy is
\be
\label{Kc}
K_c = \frac{1}{2} \sum_i m_i |{\bf v}_i|^2
\ee
where
\be
\label{vp}
{\bf v}_i = a(t) \frac{d {\bf x}_i}{dt} = \frac{d {\bf r}_i}{dt} -
H(t) {\bf r}_i \;.  
\ee 
If the system is virialized in a region of
size $R_0 \ll L$ then
\be
\left( \frac{d {\bf r}_i}{dt} \right)^2 \sim \frac{GM}{R_0} 
\gg H^2(t) {\bf r}_i^2 \approx G\rho_0R_0^2 \;.
\ee
In this approximation 
\be
\label{vp_approx}
\frac{d{ \bf r}_i}{dt} \approx {\bf v}_i 
\ee
so that  the kinetic energy in physical coordinates
(Eq.\ref{Kp}) is 
well approximated by the peculiar kinetic energy (Eq.\ref{Kc}), i.e.  
$K_p \approx  K_c$.


\subsection{The  Layzer-Irvine equation}
\label{energy_non_conservation} 

{  Let us briefly recall the expression of the LI equation.}
By defining {  the total peculiar energy as } 
$\tilde{E}=(K_c + W_p)$ the LI equation can be written as
\be 
\label{lz1}
\frac{d(a \tilde{E})}{da} = -K_c 
\;. 
\ee
Integrating Eq.\ref{lz1} we find 
\be 
\label{Alpha}
A(a) = \ \frac{ |a  \tilde{E}(a) -  a_0  \tilde{E}(a_0)|} 
{ \int_{a_0}^{a} K_c(a) da } = 1 \;.
\ee
In order to test whether the LI equation is numerically satisfied
during the time integration, one may measure the deviation of $A(a)$
from 1.  In particular, we will determine the expansion factor
$a_{LI}$ such that
\be
\label{abreak}
|A(a_{LI}) - 1| <0.1 \;\; \mbox{for} \;\; a < a_{LI} \;,
\ee
and larger than 0.1 for $a > a_{LI}$. This allows us to
quantitatively determine the range of redshifts in which the LI
equation is well approximated in the simulations, i.e. in which the LI
test is satisfied. Note that at very early times, when the initial
velocities are close (or equal) to zero the denominator of
Eq.\ref{Alpha} approaches zero and thus $A(a)$ is difficult to be
precisely determined. Otherwise for long enough times there is not any
particular numerical problem in the determination of Eq.\ref{Alpha}.




\section{Comparison between the evolution in a static and expanding background} 
\label{comparison_short_long}

We first compare the evolution of an isolated and uniform over-density
in an expanding background with periodic boundary conditions (EXP)
with the evolution of the same system in a static background and with
open boundary conditions (STAT).  We consider here the simulation M4a
(see Tab.\ref{table}), which we have chosen as a reference for the
reasons illustrated below. Results for the other simulations are
discussed in the next sections.

{Note that the difficulty of integrating the cold collapse occurs
  only around the strongest collapse phase of the system, i.e. at $t
  \approx \tau_c$, a time scale which is much smaller than any
  relevant value of the expansion factor considered in the what
  follows.}


\subsection{Short time evolution} 
\label{short_time_evolution} 

\begin{figure}
\vspace{1cm} {
\par\centering
  \resizebox*{9cm}{8cm}{\includegraphics*{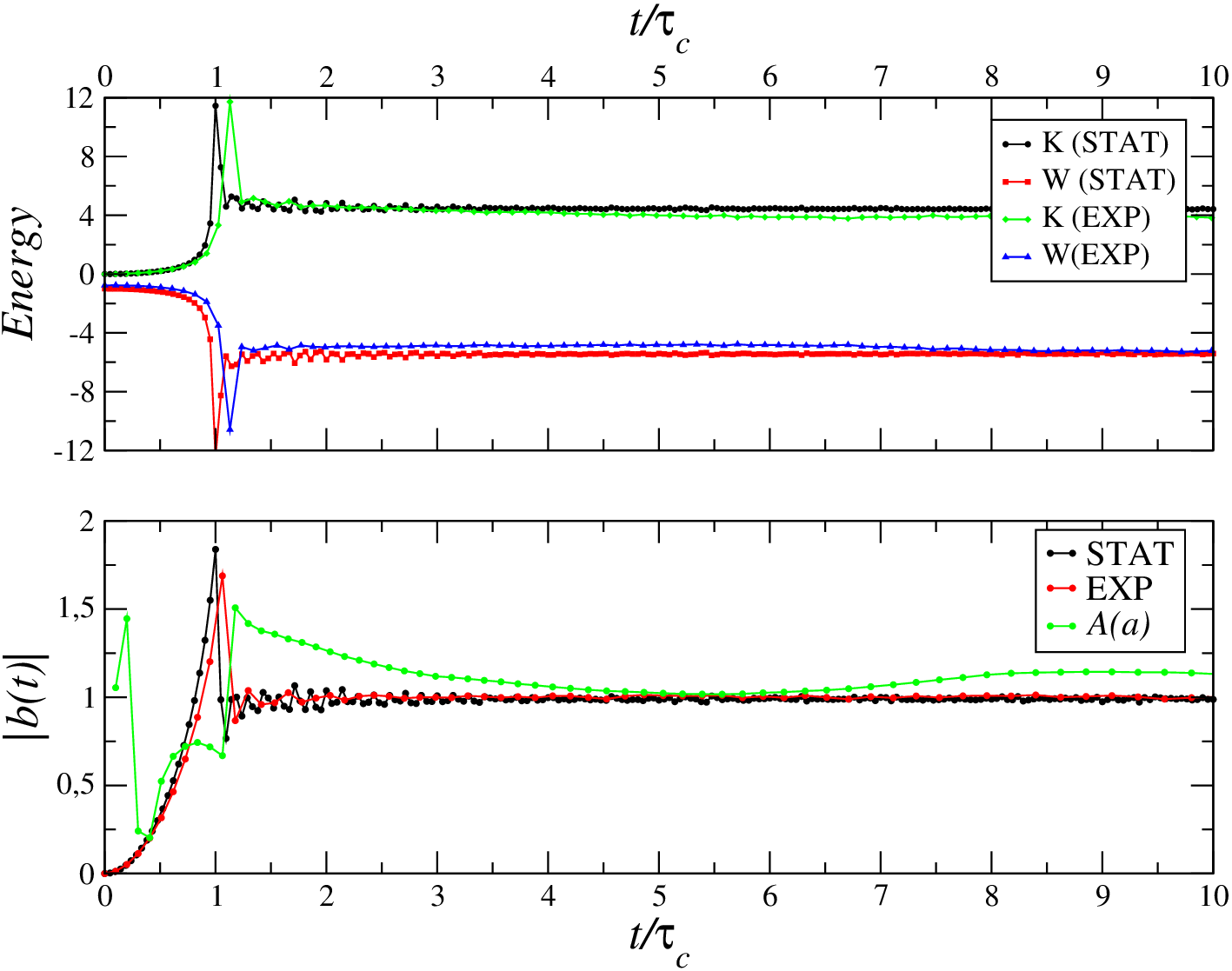}} 
\par\centering }
\caption{Upper panel: Energy in physical coordinates ($K_p$
  Eq.\ref{Kp} and $W_p$ Eq.\ref{Wp}) normalised to the initial total
  energy as a function of time (computed by using
  Eq.\ref{time-elapsed} for the EXP case), together with the
  corresponding behaviour of the same system in the STAT case.  Bottom
  panel: virial ratio $|b(t)|$, together with the LI test
  (Eq.\ref{Alpha}) for the EXP case.}
\label{Energy_OBC_EXP}
\end{figure}

Fig.\ref{Energy_OBC_EXP} (upper panel) shows the potential and kinetic
energy in physical coordinates (normalised to the initial potential
energy $W_0$). As for the STAT case the potential (kinetic) energy
decreases (increases) to reach its minimum (maximum) at $t\approx
\tau_c$, i.e. at the strongest phase of the collapse. For $t>\tau_c$
the bounded structure is stationary, i.e. $W_p \approx$ const., and it
satisfies the virial equilibrium (Fig.\ref{Energy_OBC_EXP} --- bottom
panel). A fraction of about $f_p \approx 25\%$ of the particles gain
kinetic energy so that, after the collapse, they are ejected from the
system moving as free particles thereafter.

Energy conservation for the STAT case can easily be controlled to a precision
better than $\approx 0.01 \%$ 
(see, e.g., \citet{aarseth} and references therein).  From the LI test
(Fig.\ref{Energy_OBC_EXP} --- bottom  panel) in the EXP case we
find $A(t) \approx 1$ with relatively large fluctuations up to the
maximum contraction phase. However, these fluctuations are damped
after the collapse: we thus conclude that in this range of time the
system is still in the regime $a<a_{LI}$ (see Eq.\ref{abreak}).

\begin{figure}
\vspace{1cm} { \par\centering
  \resizebox*{9cm}{8cm}{\includegraphics*{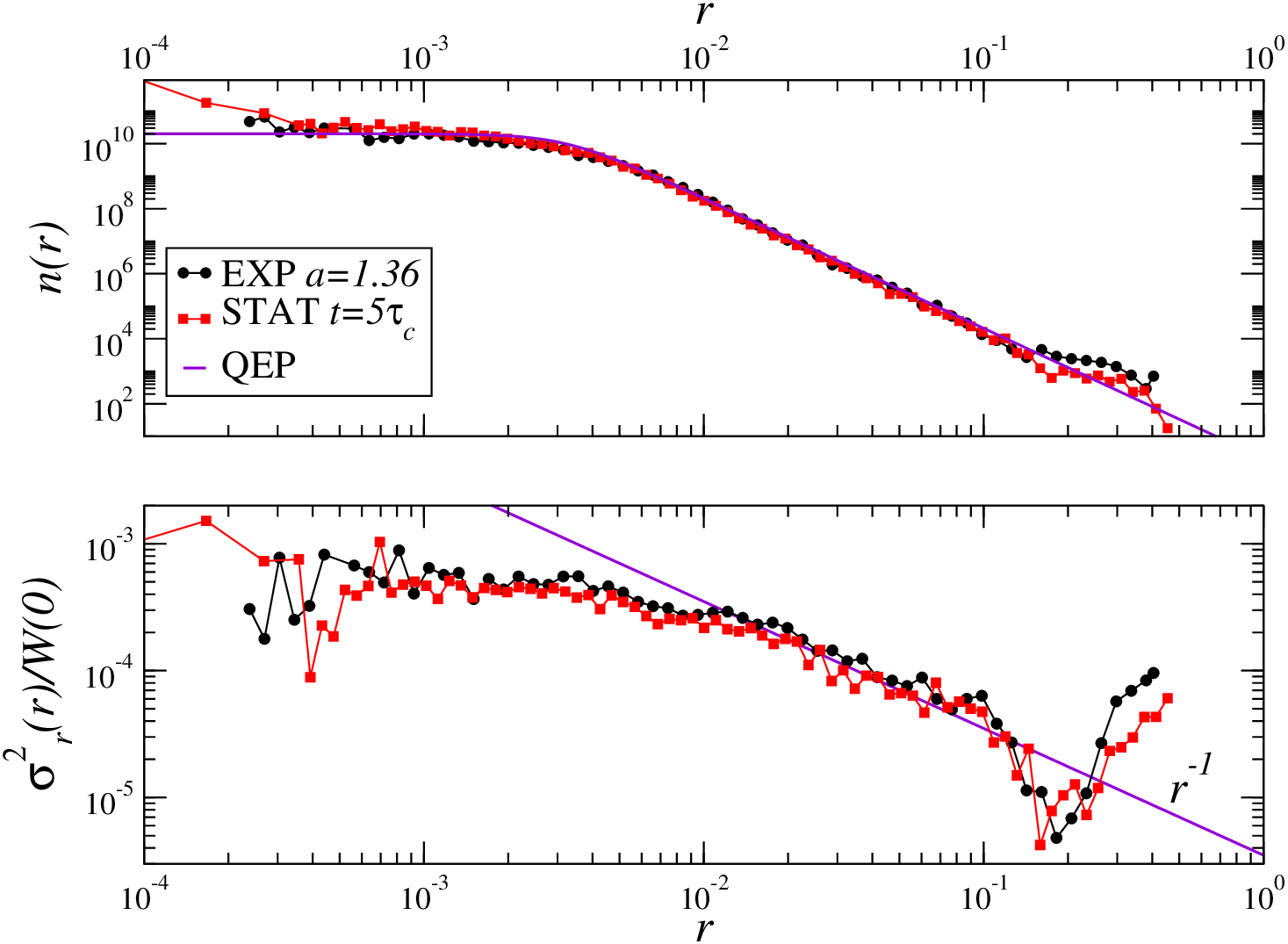}}
  \par\centering }
\caption{Comparison of the density (upper panel) and velocity (bottom
  panel) profiles of the same system evolved in an expanding
  background (EXP at $a=1.36$ --- black circles) and in a static
  background with open boundary conditions (STAT at $t=5 \tau_c$ ---
  red squares).  The best fit with Eq.\ref{nrb0} (QEP) and a line with
  slope $-1$ are shown as references (see Eq.\ref{mainb2}).  }
\label{DP_OBC_EXP}
\end{figure}

The density profile (Fig.\ref{DP_OBC_EXP} --- upper panel) is well
fitted by Eq.\ref{nrb0} both in the EXP and STAT case. The radial
velocity dispersion (Fig.\ref{DP_OBC_EXP} --- bottom panel) {  shows
  the Keplerian behaviour described by Eq.\ref{mainb2} Therefore the
  pseudo phase-space density obeys to Eq.\ref{mainb3}, i.e. it decays
  as $\Phi (r) \sim r^{-5/2}$.  }




\subsection{Long time evolution} 
\label{long_time_evolution} 

The situation discussed in Sect.\ref{short_time_evolution} drastically
changes at longer times in the EXP simulation.  Indeed, already for
moderate values of the expansion factor, the evolution of the system
rapidly deviates from the short-time behaviour, and correspondingly
the properties of the QSS become different from
Eqs.\ref{mainb1}-\ref{mainb3}.

\begin{figure}
\vspace{1cm}
{
\par\centering \resizebox*{9cm}{8cm}{\includegraphics*{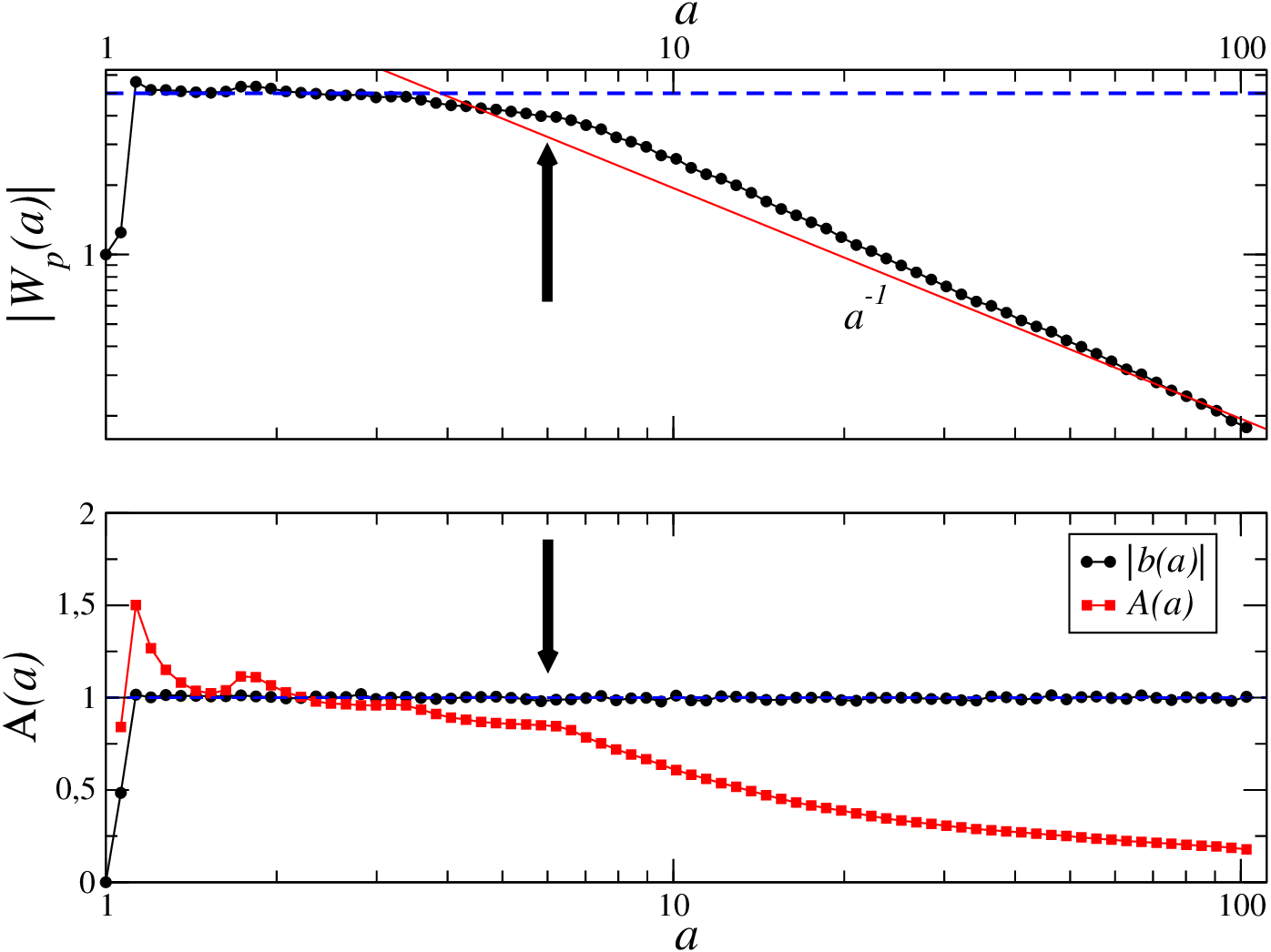}}
\par\centering
}
\caption{Upper panel: Absolute value of the potential energy in
  physical coordinates $W_p(a)$ (Eq.\ref{Wp}) as a function of the
  scale factor.  The approximate value of $a_{LI}$, separating the
  regime where $W_p \approx $ const. from the one where $|W_p| \sim
  a^{-1}$, is reported as reference.  
Bottom  panel:   LI test
  (Eq.\ref{Alpha}) and virial ratio (absolute value) for particles
  with negative energy.}
\label{Energy_EXP_LONG}
\end{figure}

\begin{figure}
\vspace{1cm}
{
\par\centering \resizebox*{9cm}{8cm}{\includegraphics*{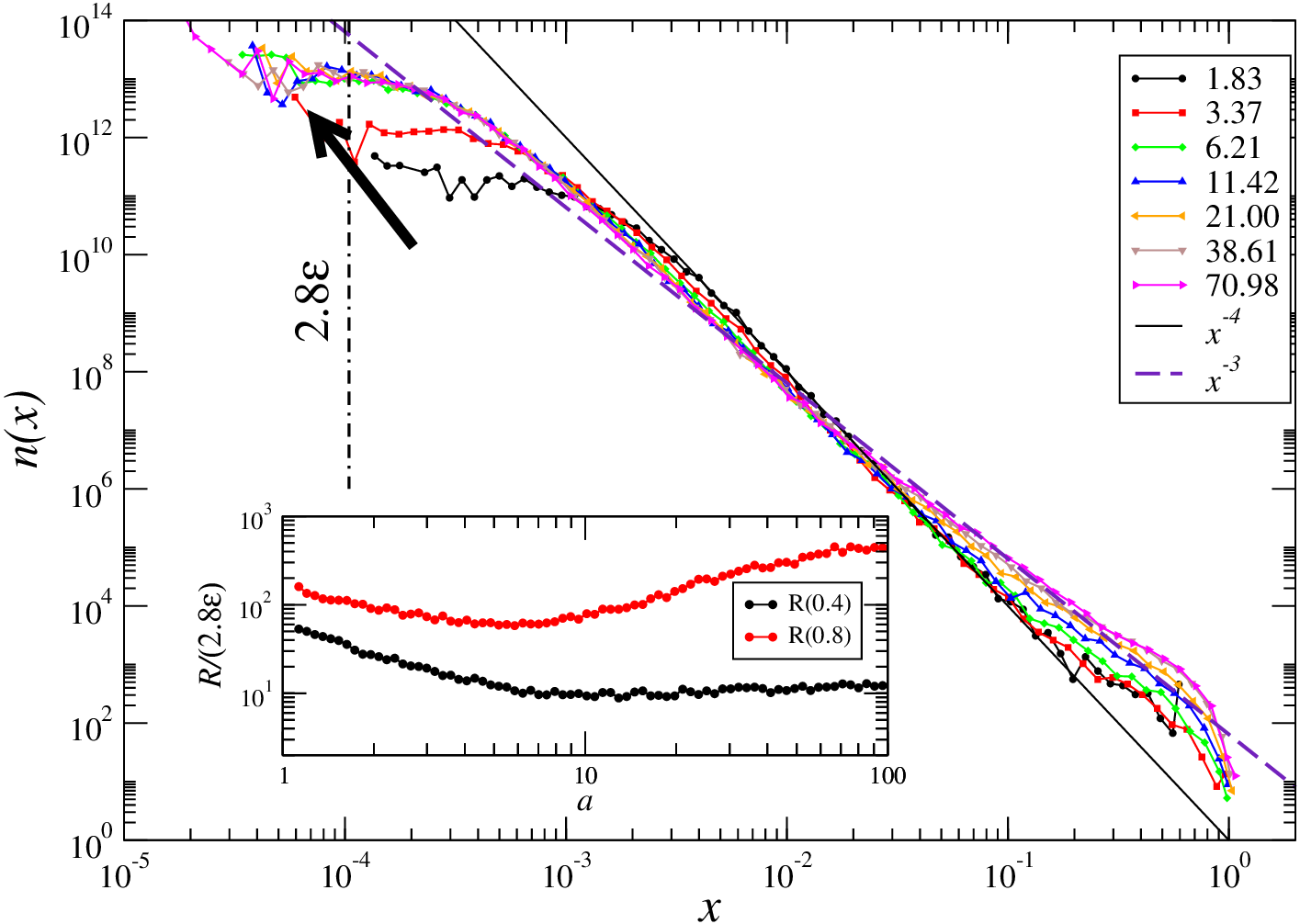}}
\par\centering
}
\caption{Evolution of the density profile in comoving coordinates.
  Two reference lines with slope -4 and -3 are reported as a
  reference. The arrow shows the direction of the density
  profile evolution.  Inset panel: behaviours in function of the scale
  factor of the length scale $R(0.4)$ ($R(0.8)$), including the $40\%$
  ($80\%$)of the mass of the virialized structure, normalised to $2.8
  \varepsilon$. }
\label{DP_CC_T2}
\end{figure}

We find that $W_p \approx$ const., i.e. the structure is in a QSS and
in the stable clustering regime, only for $a< a_{LI} \approx 6$ (see
Fig.\ref{Energy_EXP_LONG}).  Instead, for $a> a_{LI}$ we find that
$|W_p| \sim 1/a$ (and thus $W_c \approx$ const). Thus the breakdown of
the LI test corresponds to a departure from the stable clustering
regime.
Indeed, the density profile in comoving coordinates for $a> a_{LI} $
stabilises to an almost ``asymptotic'' shape which does not change
anymore at small radii and slowly changes at large ones (see
Fig.\ref{DP_CC_T2}). This is characterised by an approximate $x^{-3}$
decay extending over about three decades, i.e. from $\sim 10^{-3} L$
to $L$.

Note that the softening length is fixed in comoving coordinates and
thus in physical coordinates it increases as $\varepsilon_{p} =
a\varepsilon$. When the structure is in the stable clustering regime,
its size in physical coordinates $R_{p}$ is constant, and thus
$R_{p}/\varepsilon_{p} \propto 1/a$: if the stable clustering regime
persists long enough then the size of the structure will become of the
order of, or even smaller than, the softening length.  However, we
noticed that beyond a certain time the structure becomes stable in
comoving coordinate and thus its size in physical coordinates scales
as $R^*_{p} = a R^*_{cc}$, where $R^*_{cc}$ is the constant size in
comoving coordinates.  Therefore for $a>a_{LI}$ one approximately
finds that $R^*_{p}/\varepsilon_{p} \approx R^*_{cc}/\varepsilon
\approx const.$

The length scale $R^*_{cc}$ can be estimated, for instance, by
measuring the radius $R(0.4)$ ($R(0.8)$) including the 40\% (80\%) of
the mass of the virialized structure. As shown in inset panel of
Fig.\ref{DP_CC_T2}, for $a>a_{LI}$, $R(0.4)$ and $R(0.8)$ remain
respectively larger than 10 and 100 times the scale beyond which the
force is exactly Newtonian, i.e. 2.8$\varepsilon$. In this condition
the virial ratio fluctuates around -1 at all times (see
Fig.\ref{Energy_EXP_LONG}). This implies that the Newtonian value of
the mean field potential is well approximated by the smoothed
potential and thus the softening length does not perturb the mean
field dynamics. (Note that this result is confirmed by the behaviours
measured by using other values of the softening length, as well as by
an analysis of the gravitational potential energy in function of
$\varepsilon$ --- see discussion in Sect.\ref{Softening_length}).

\begin{figure}
\vspace{1cm}
{
\par\centering \resizebox*{9cm}{8cm}{\includegraphics*{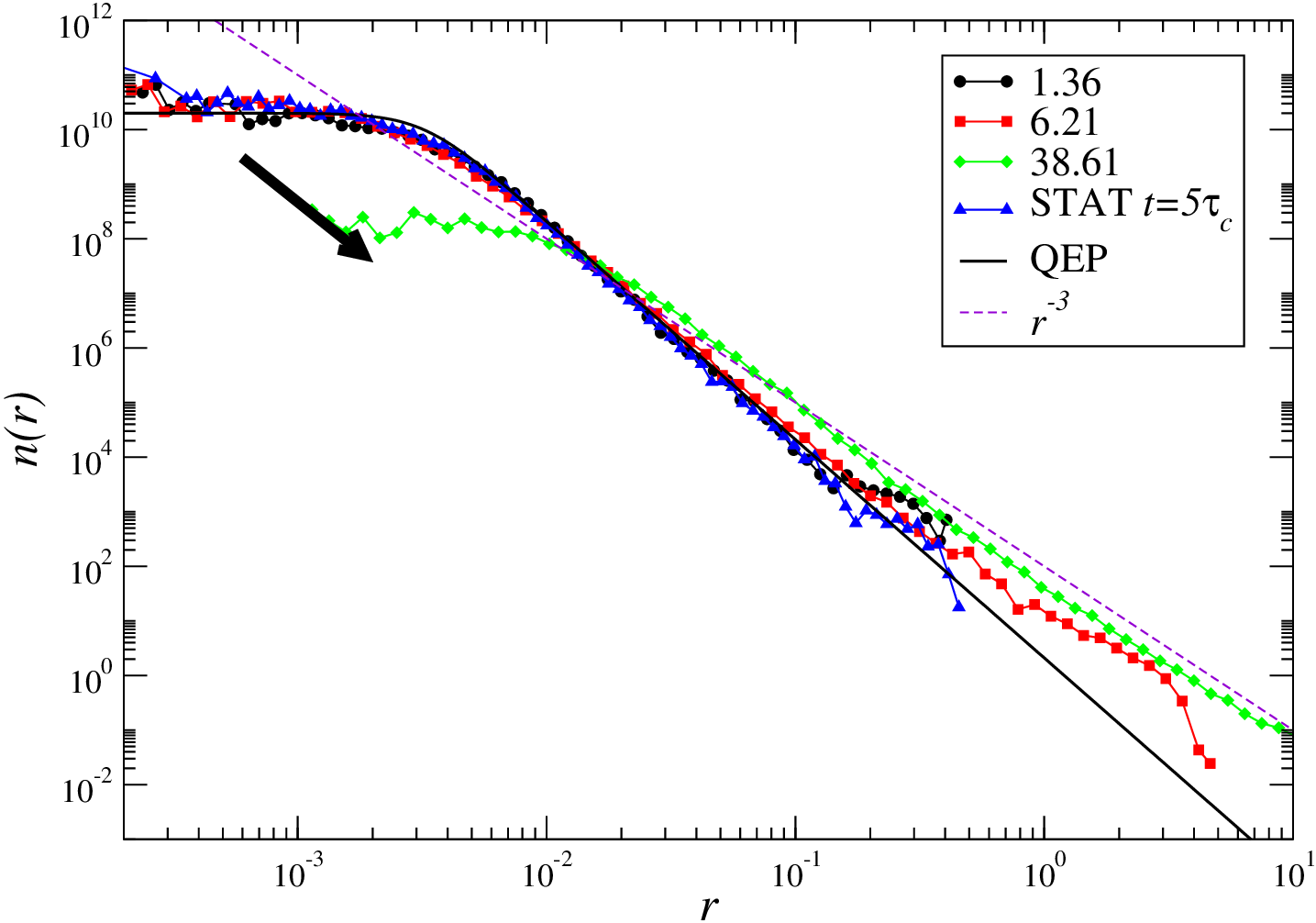}}
\par\centering
}
\caption{Density profile in physical coordinates for the simulation
  M4a at different scale factors $a$ (see labels) together with the
  density profile for the STAT case.  The arrow shows the direction of
  the density profile evolution in the cosmological case.  Further it
  is shown the best fit given by Eq.\ref{nrb0} (QEP) and a reference
  line with slope -3.}
\label{dp_T2a_1}
\end{figure}

Coherently with the behaviour of the energy, the density profile in
physical coordinates (see Fig.\ref{dp_T2a_1}) does not change for $a <
a_{LI}$, and it is well fitted by Eq.\ref{nrb0}: this situation
corresponds to the QSS in the stable clustering regime.  Note that, as
time passes, the spatial extension of the halo, i.e. the range of
radii where $n(r) \sim r^{-4}$, increases.  Indeed, as time passes
particles with energy $e_p \ltapprox 0$ may go asymptotically far from
the structure core \citep{sl12a}.

On the other hand, for $a> a_{LI}$, we find that $n(r)$ continues to
evolve: at large radii it decays as $\sim r^{-3}$ and at small radii
the density does not flatten as for the STAT case. { A \citet{nfw}
  (NFW) profile

  \be
\label{nfw_eq}
n(r) = \frac{n_c}{(r/r_c)(1+(r/r_c))^2}\;.  
\ee 
gives} a reasonable
fit for $a> a_{LI}$ (bottom panels of Fig.\ref{dp_T2a_1}): in
particular the large scale tail shows a well-defined $r^{-3}$
decay. Therefore only when stability {in physical coordinates} is
observed, the density profile is consistent with the expected one,
i.e.  Eq.\ref{nrb0}; when stability { in physical coordinates} is
broken then the density profile changes its functional behaviour
approaching the NFW one.

Similarly, the radial velocity dispersion profile in physical
coordinates (Fig.\ref{sigma2_T2}) shows an approximated Keplerian
decay at large scale as in the STAT case.  However, while for $a <
a_{LI}$ the different curves collapse on each other, implying that the
virialized structure is in the stable clustering regime, for $a >
a_{LI}$ {  the velocity profile marks as well } a breaking of the
stable clustering regime. In particular, the slope remains $-1$,
i.e. halo particles are still orbiting in a central gravitational
potential but with an {\it amplitude} that decreases with time.  {This
  is consistent with the fact that the potential energy in physical
  coordinates $W_p$ increases with the scale factor for $a > a_{LI}$
  (as $W_c \approx$ const.): as time passes $R^*_{p} \propto a$ and
  thus $W_p$ increases proportionally to $-1/a$.
\begin{figure}
\vspace{1cm}
{
\par\centering \resizebox*{9cm}{8cm}{\includegraphics*{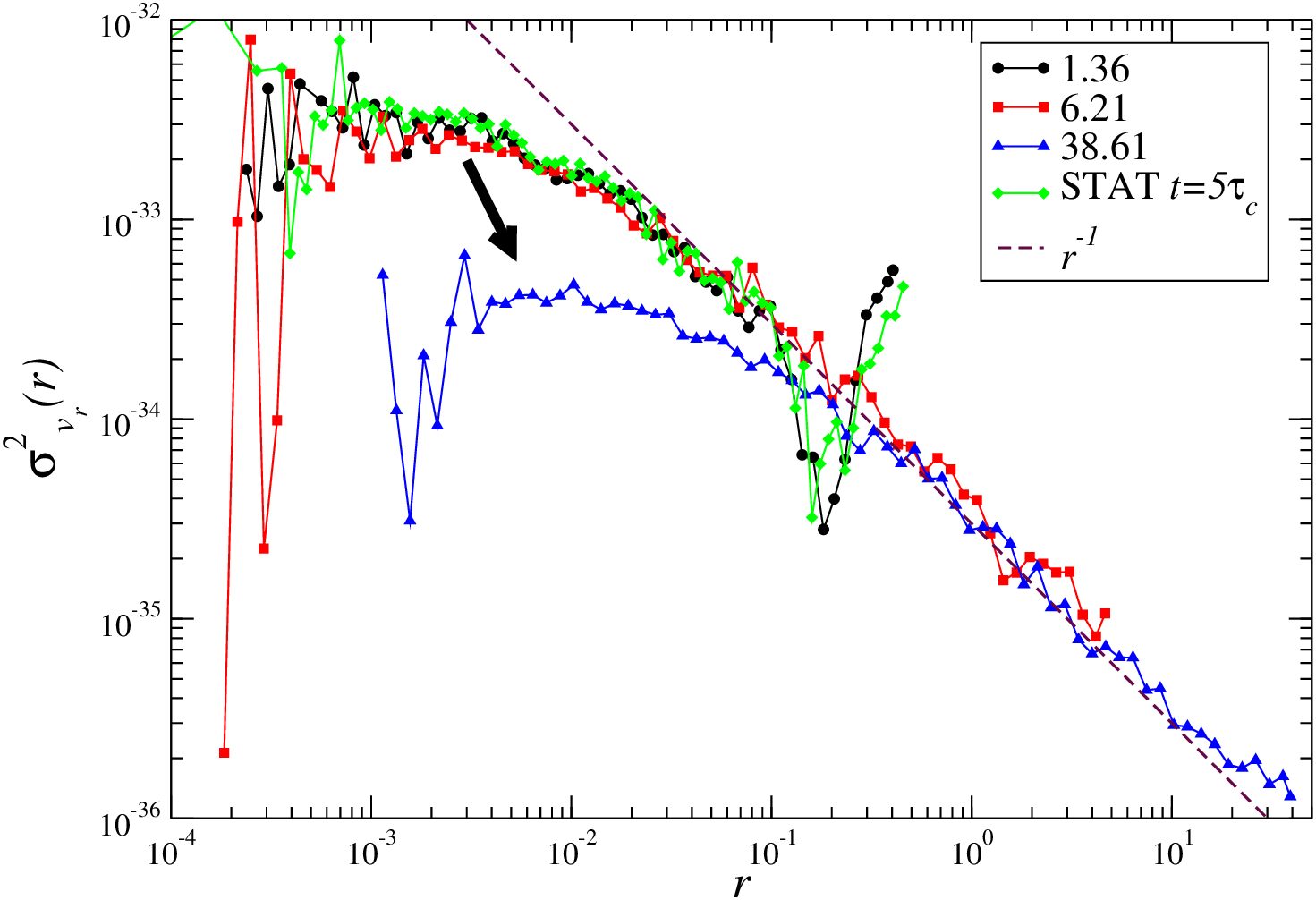}}
\par\centering
}
\caption{Radial velocity dispersion profile in physical coordinates at
  different scale factors $a$ (see labels) together with the profile
  for the STAT case. The arrow shows the direction of the density
  profile evolution in the cosmological case. A line with slope
  $r^{-1}$ (see Eq.\ref{mainb2}) is reported as reference.}
\label{sigma2_T2}
\end{figure} 
 {\it Thus for $a > a_{LI}$ there is an energy injection into the
   system which causes an increase of the total energy}: this occurs
 slowly enough that the system can rearrange its phase space
 configuration to satisfy the virial theorem (see the bottom panel of
 Fig.\ref{Energy_EXP_LONG}). Therefore the energy increase corresponds
 to decrease of the kinetic energy: as $b \approx -1$ we have $\Delta
 E = - \Delta K_p$.  Note the physical size of the structure increases
 almost simultaneously with the violation of the LI equation: this
 implies that the energy injection is spurious as otherwise the LI
 test should be verified.}

\begin{figure}
\vspace{1cm} \par\centering
\resizebox*{9cm}{8cm}{\includegraphics*{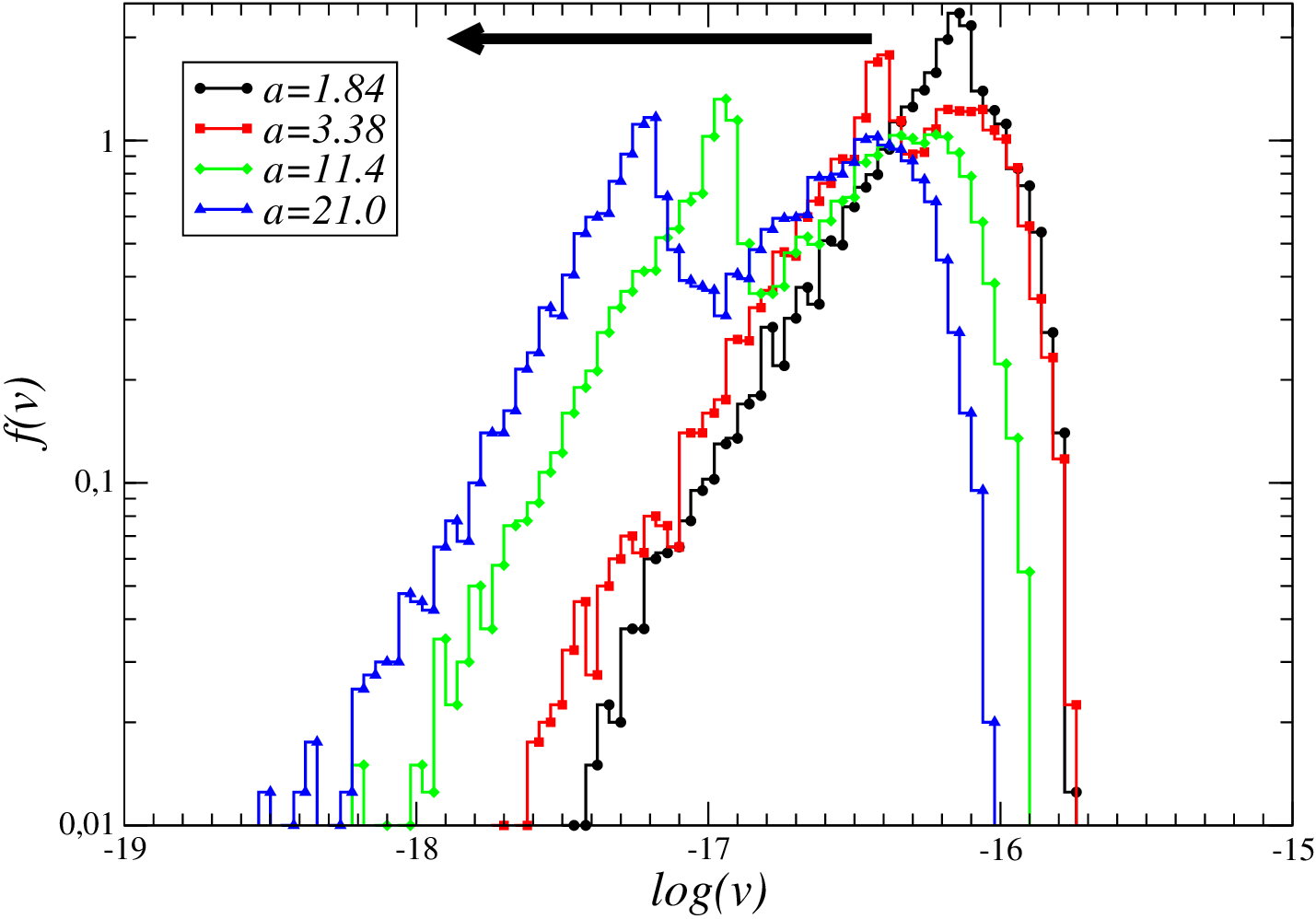}} \par\centering
\caption{Evolution of the probability distribution
  function (PDF) of the absolute value of the velocity at different
  scale factors (see labels). The arrow shows the direction 
of the evolution.}
\label{fv_T2_A}
\end{figure}

The decrease of the kinetic energy corresponds to the fact that high
velocity particles are slowed down. This is shown in Fig.\ref{fv_T2_A}
(upper panel) where it is reported the evolution of the probability
distribution function (PDF) $f(v)$ of the absolute value of the
velocity.  While for $a < a_{LI}$ we find that $f(v)$ remains almost
stable, for $a > a_{LI}$ high velocity particles are slowed down and
correspondingly the high velocity tail of $f(v)$ is significantly
reduced.  {  However the shape of $f(v)$ remains almost the same for
  $a > a_{LI}$.}  The failure of the LI test and consequent breakdown
of the stable clustering regime is thus induced by the difficulty of
integrating high velocity particles.

 {  High velocities are typically associated with close scatterings
   and the} limited ability to resolve hard scattering between close
 neighbours, through which particles can increase their velocity, can
 also be traced by looking at the evolution of the nearest neighbour
 distribution $\omega(x)$ in comoving coordinates (see
 Fig.\ref{fv_T2_A}): indeed, one may note that while for $a<a_{LI}$
 $\omega(x)$ shows a peak that is shifted at smaller and smaller
 scale, corresponding to the fact that the structure is stable in
 physical coordinates, for $a>a_{LI}$ $\omega(x)$ does not change
 anymore reaching an ``asymptotic'' shape. This is a clear effect of
 the limited small scale resolution of the numerical code.
\begin{figure}
\vspace{1cm}
\par\centering \resizebox*{9cm}{8cm}{\includegraphics*{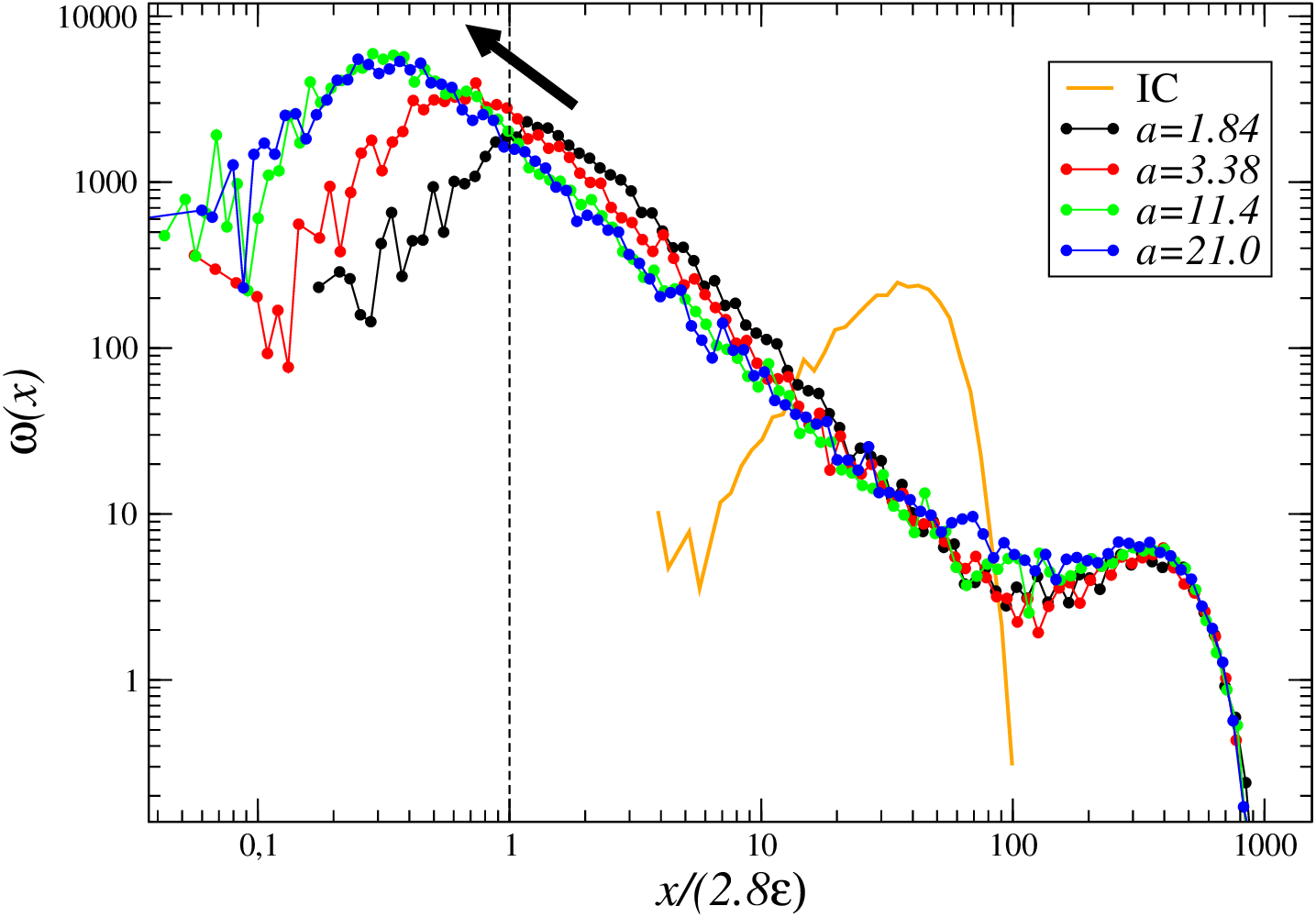}}
\par\centering
\caption{Evolution of the nearest neighbour distribution in comoving
  coordinates at different scale factors (see labels).  The arrow
  shows the direction of the evolution. The initial distribution (IC)
  is also shown for comparison.}
\label{fv_T2_b}
\end{figure} 

Finally we note that from the behaviours of the density profile and of
the radial velocity dispersion, we can easily find that the pseudo
phase-space density, for $a< a_{LI}$ and for $r>r_c$, scales as
$\Phi(r) \sim r^{-5/2}$, i.e., the behaviour expected for the QEP (see
Eq.\ref{mainb3}).  {  On the other hand, for $a> a_{LI}$, at large
  radii, we find $\Phi(r) \sim r^{-3} \times r^{3/2} \sim r^{-1.5}$.}


\subsection{Summary} 

Let us summarise the situation in the two regimes that we have
identified., i.e.  $a<a_{LI}$ and $a>a_{LI}$.

\begin{enumerate} 

\item For $a < a_{LI}$ the structure is in a QSS (i.e., $|W_p|
  \approx$ const.) and the density and velocity profiles are described
  by the collision-less equilibrium behaviours
  (Eqs.\ref{mainb1}-\ref{mainb3}).  This situation corresponds to the
  stable clustering regime: the physical size and shape of the
  structure remains invariant during time evolution, modulo a marginal
  spatial extension of the power-law tail due to high velocity, but
  bounded, particles.

\item {For $a > a_{LI}$ the structure is (almost) frozen in
  comoving coordinates and therefore its size in physical coordinates
  increases and its potential energy tends to zero as $W_p \propto
  -1/a$.  The breakdown of the IL test marks a problem in the
  numerical integration of the particles equations of motion in this
  regime: this occurs because there is an injection of energy which
  causes the expansion of the structure in physical space.}
  Correspondingly the density profile develops a $r^{-3}$ tail, {
    approaching a} NFW profile.  The velocity dispersion profile still
  decays in a Keplerian way but with an amplitude that decreases with
  time equilibrating the smaller potential energy (in absolute value)
  of the structure, so that the virial ratio remains $b \approx -1$.
\end{enumerate}

The decrease of the system kinetic energy for $a > a_{LI}$ corresponds
to the fact that high velocity particles are slowed down and the PDF
of the particles velocity $f(v)$ shifts toward smaller velocity
values.  This is indeed the consequence of the inaccurate integration
of two-body collisions: it becomes more difficult to numerically
resolve high velocity particles.  As shown by the behaviour of the
nearest neighbour distribution $\omega(x)$ these high velocity
particles are also close neighbours. Indeed, $\omega(x)$ does not
evolve anymore for $a>a_{LI}$, as it should if stable clustering were
satisfied.  We thus conclude that the inaccurate integration of hard
collisions induces the change of regime at $a_{LI}$.

Note that $a_{LI} \approx 6$ corresponds to a time (see
Eq.\ref{time-elapsed}) $(t_{break}-t_0)/\tau_c \approx 50$ that is
much smaller than the collisional relaxation time scale (see
Eq.\ref{tau2}) $\tau_2 \approx 10^4 \tau_c$ for a system with $N=10^4$
particles.  This implies that the break-down of the QSS is not related
to the expected evaporation due to two-body collisions for $t>\tau_c$.

Finally we note that the breakdown of the QSS, if physically
originated, should not be associated with the breakdown of the LI test
as we find. This shows that such a breakdown is induced by spurious
reasons that we identify in the role of the inaccurate integration of
high velocities particles and thus of hard scatterings. We will return
on this important point in Sect.\ref{Softening_length}.


\section{Role of numerical integration parameters} 

\label{numerical_effects}

In this section we study the effect of varying some of the most
relevant numerical parameters that characterise the numerical
integration: the softening length $\varepsilon$, the parameter that
controls the time-stepping accuracy $\eta$, the opening angle $\theta$
of the tree algorithm and the accuracy of the relative cell-opening
criterion $\phi$ \footnote{i.e., the parameter denominated 
  {\tt ErrTolForceAcc} in the code}.


\subsection{Softening length}
\label{Softening_length}

The softening length~\footnote{recall that $\varepsilon$ is given in
  units of the box side $L$} has been varied in a range such that: (i)
$\varepsilon < R_0$, i.e. it is smaller than the initial size of the
structure so that the gravitational mean field force is well
approximated at least at early times and (ii) the lower limit to
$\varepsilon$ is due to numerical constraints: $\varepsilon > 10^{-6}$
otherwise the integration time rapidly becomes too long and the
simulation becomes unfeasible.

\begin{figure}
\vspace{1cm} { \par\centering
  \resizebox*{9cm}{8cm}{\includegraphics*{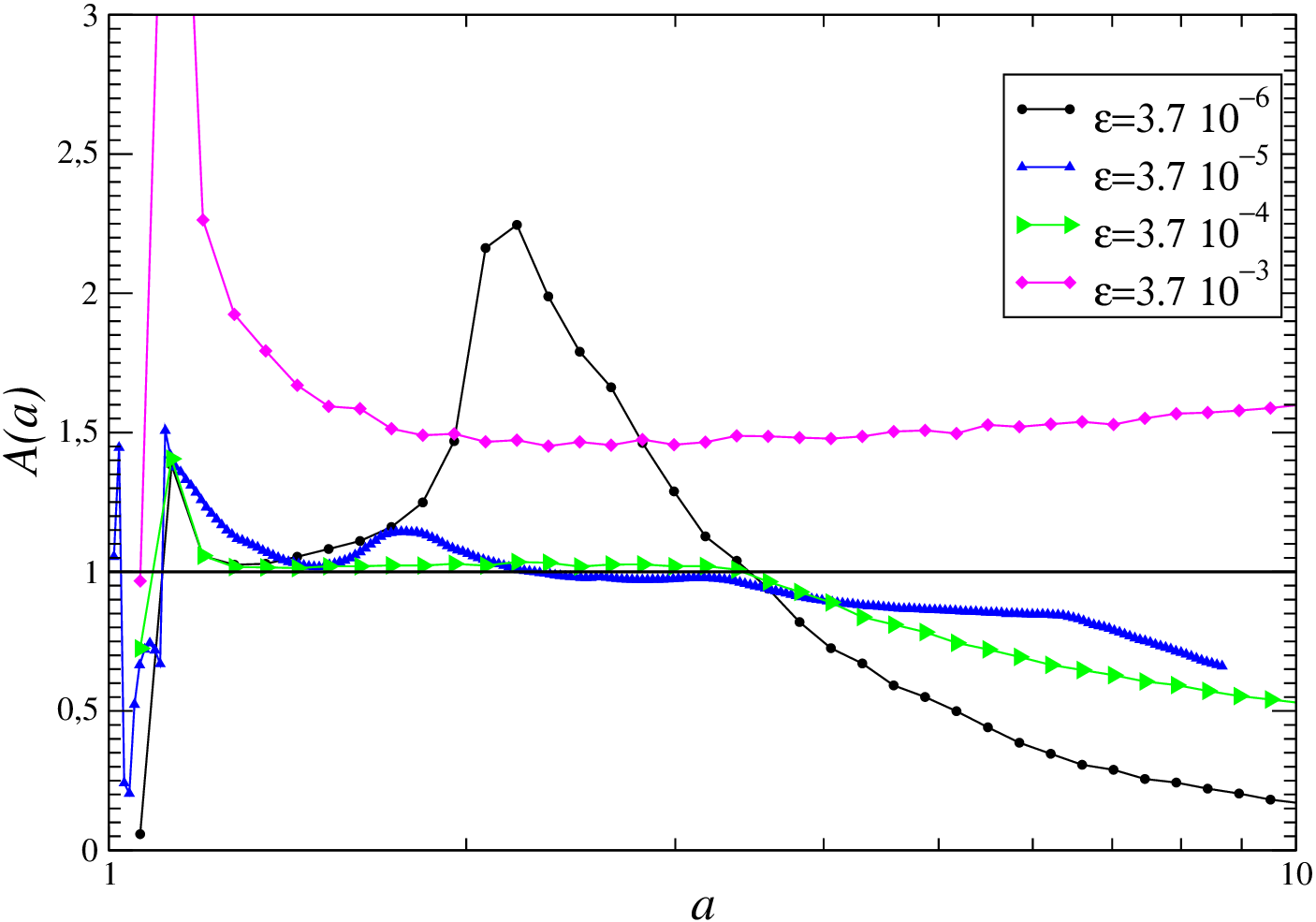}} \par\centering }
\caption{Behaviour
  of $A(a)$ (Eq.\ref{Alpha}) for  simulations with different
  $\varepsilon$ (see Tab.\ref{table} for details).}
\label{Energy_epsilon} 
\end{figure}

The potential energy in physical coordinates $W_p$ and the IL test
(Fig.\ref{Energy_epsilon}) for simulations of the same system with
$N=10^4$ particles and with different values of $\varepsilon$ (see
Tab.\ref{table}) --- changed by a factor $10^4$ --- show behaviours
coherent with those discussed in Sect.\ref{long_time_evolution}.
Namely the two regimes found above, corresponding to $W_p \approx$
const. for $a < a_{LI}$ and $|W_p| \sim a^{-1}$ for $a> a_{LI}$, still
characterise the temporal evolution. However, {\it the transition
  between them depends on the softening length}, i.e., $a_{LI} =
a_{LI}(\varepsilon)$.

In particular, we find that $a_{LI} \approx 6$ for $\varepsilon \in
[3.7 \times 10^{-5},3.7 \times 10^{-4}$], while outside this range
$a_{LI} \approx 1$.  This result is in agreement with \citet{jsl13}
who concluded that at most in that narrow range of $\varepsilon$ can
the behaviour required, a QSS in the stable clustering regime, be
reproduced by the N-body method.

In order to understand the effect of a softening length which is too
large with respect to the size of the structure, we have computed the
gravitational potential energy of the structure with density profile
described by Eq.\ref{nrb0}, in the case in which the pair potential is
the {\tt GADGET} one with different values $\varepsilon$.  {  We
  find that} the gravitational potential energy is very well
approximated by the Newtonian value when $R(0.4)/(2\varepsilon)
\approx 10$ and/or $R(0.8)/(2\varepsilon) \approx 100$, where $R(0.4)$
and $R(0.8)$ are the radii in comoving coordinates including
respectively the 40\% and 80\% of the mass of the virialized
structure.

\begin{figure}
\vspace{1cm} { \par\centering
  \resizebox*{9cm}{8cm}{\includegraphics*{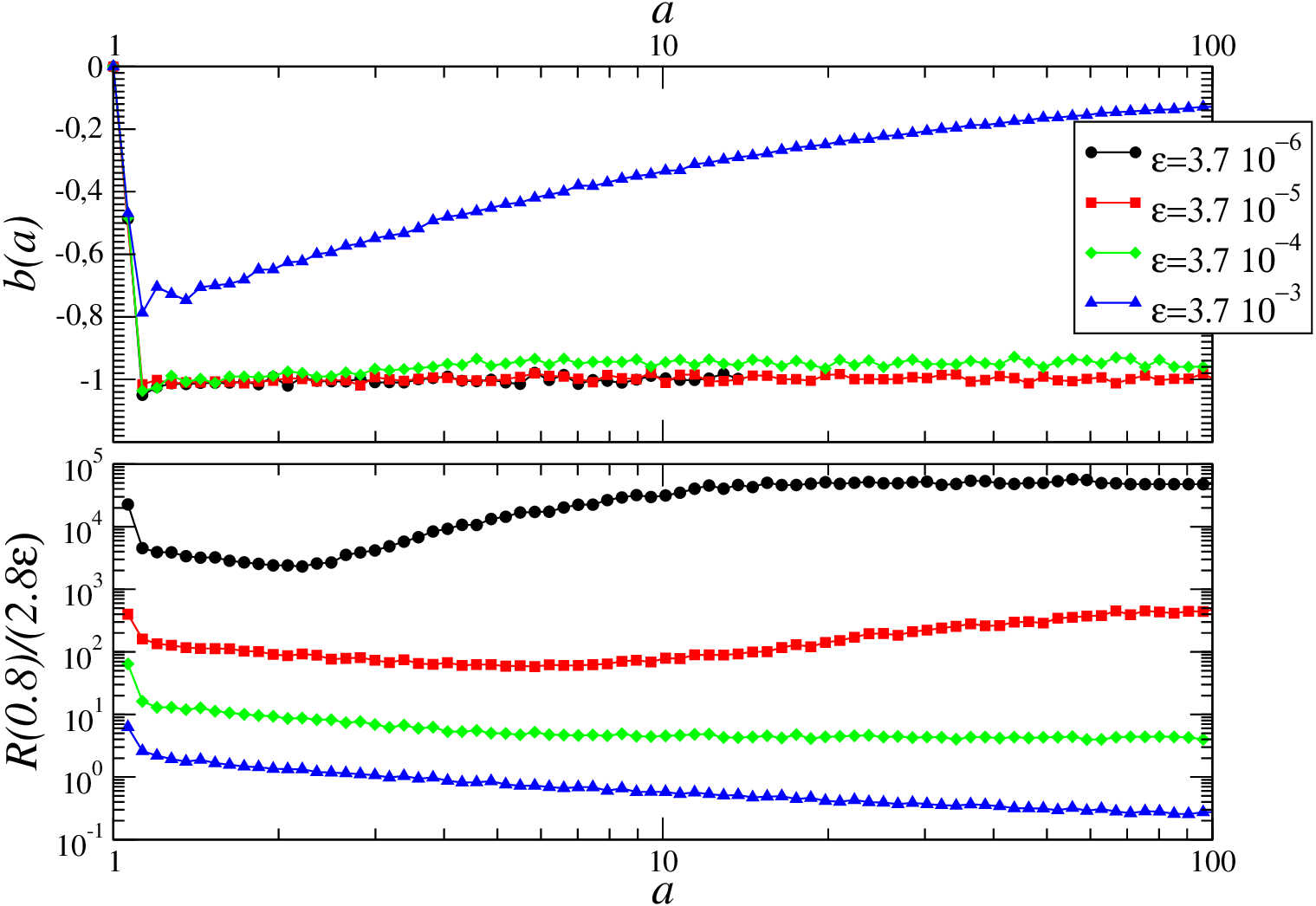}} \par\centering }
\caption{Upper panel: virial ratio as a function of the scale factor
  in simulations with different $\varepsilon$.  Bottom panel: Length
  scale $R(0.8)$, including the $80\%$ of the mass of the virialized
  structure, normalised to $2.8\varepsilon$.}
\label{Fig10ter} 
\end{figure}

The behaviours of $R(0.8)/(2\varepsilon)$ as a function of the
expansion factor and for different values of the softening length are
reported in Fig.\ref{Fig10ter} (see the bottom panel).  For
$\varepsilon \ge 3.7 \times 10^{-4}$ we have that $R(0.4)/(2.8
\varepsilon) < 10$: in this situation the effective mean field force
due to all particles is different from Newtonian, i.e. the smoothening
length is too large with respect to the size of the system.  The
effect of a too large softening thus corresponds to a deviation of the
virial ratio from -1 (see the upper panel of Fig.\ref{Fig10ter}).

Instead, for $\varepsilon \le 3.7 \times 10^{-5}$ we find that,
because the system remains stable in comoving coordinates $R(0.8)/(2.8
\varepsilon)$ is larger than 100 for all values of the scale factor
considered, and thus the virial ratio fluctuates around -1 at all
times. Therefore we conclude that the relevant characteristic size of
the structure remains large enough with respect to the force softening
and that in this situation the Newtonian value of the mean field
potential is well approximated by the smoothed potential.

The behaviour of the density profile is also coherent with what we
have already found in Sect.\ref{comparison_short_long}: for $a <
a_{LI}$, the density profile obeys to Eq.\ref{nrb0}.
\begin{figure}
{ \par\centering \resizebox*{9cm}{8cm}{\includegraphics*{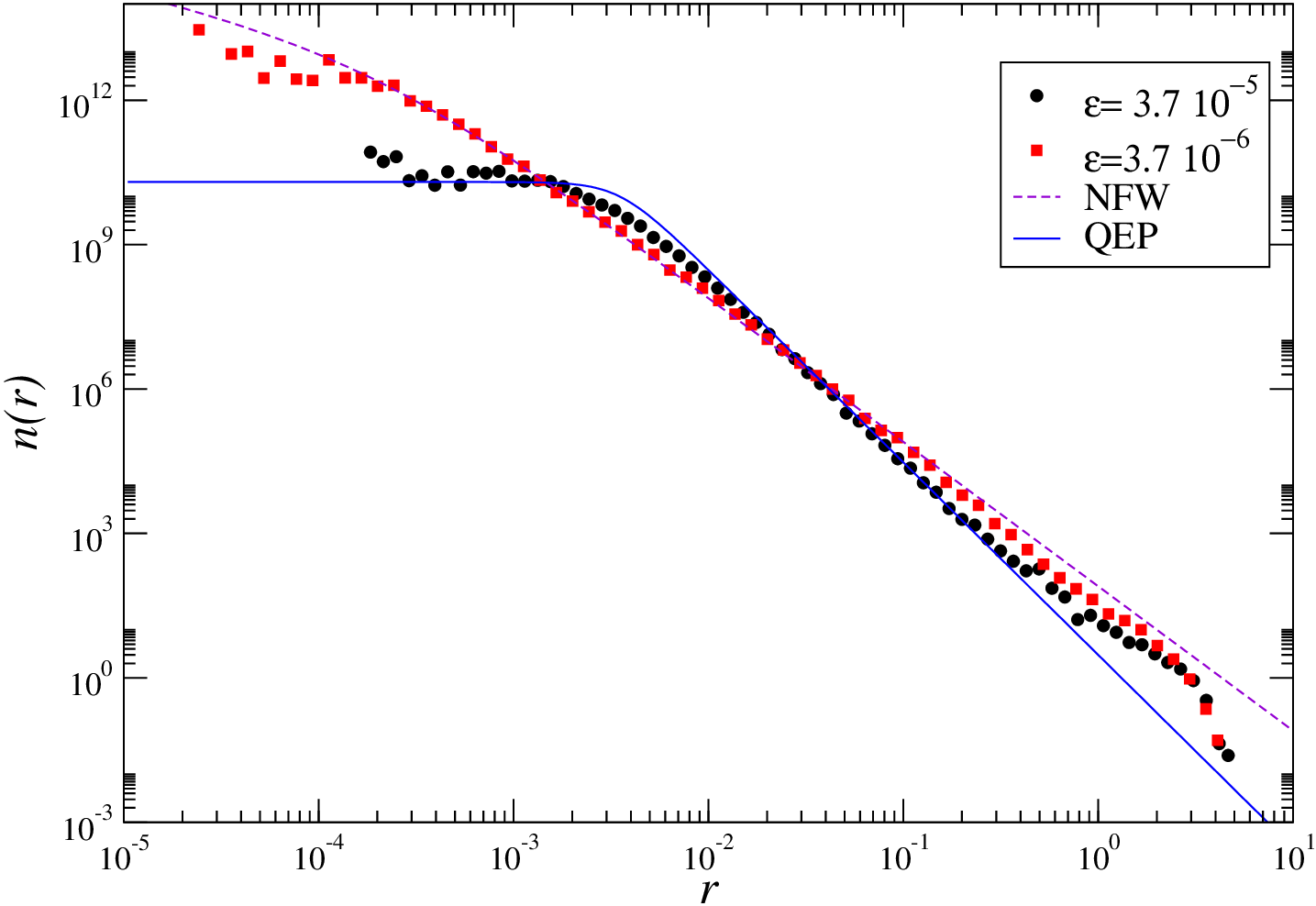}}
  \par\centering }
\caption{ Density profile in physical coordinates at $a=4.6$ for the
  simulations M4a and M1a. The best fit with the NFW and QEP profiles
  are also reported.}
\label{dp_epsilon_series_1}
\end{figure} 

 Then, for $a > a_{LI}$, we observe a breaking of the stable
 clustering regime, and the density profile rapidly changes shape.  In
 particular, it firstly deviates at small radii, displaying a
 power-law behaviour, and then at large ones where the slope tends to
 $-3$. The whole density profile can be fitted by a NFW behaviour
(see  Fig.\ref{dp_epsilon_series_1}).

{  The evolution of the the velocity PDF $f(v)$ and the nearest
  neighbours distribution $\omega(x)$ for different values of
  $\varepsilon$ are coherent with the behaviours previously
  discussed. Namely, for large enough scale factors, i.e. $a \ge 10$,
  the simulation which behaves better according to the LI test (i.e.,
  $\varepsilon= 3.7 \times 10^{-5}$) shows a more extended tail of
  $f(v)$ at high velocities than the others. This means a better
  ability of the code to follow high velocity particles.
  Correspondingly also $\omega(x)$ is peaked at smaller separations
  i.e.  close particles can be better resolved for that value of the
  softening length.  Both these facts point toward a better spatial
  (i.e., close neighbours) and temporal (i.e., high velocity
  particles) resolution when $\varepsilon= 3.7 \times 10^{-5}$.  }

{  Similarly to the behaviour of the density profile, whose small
  scale properties depend on $\varepsilon$ the nearest neighbour
  distribution and its peak shown an $\varepsilon$ dependence implying
  that the small scale properties of the system are determined by the
  spatial resolution used in the simulation.}


\subsection{Time-stepping accuracy}
\label{time_step_accuracy}

The accuracy of the time-stepping in the {\tt GADGET} code is controlled
by the parameter $\eta$: in particular, the time-stepping is fixed by
\be
\label{timestepacc} 
\Delta t = \sqrt{\frac{2 \eta \varepsilon}{\mathcal{A}_{acc}}} \;,
\ee
where $\mathcal{A}_{acc}$ is the particle acceleration.  It thus depends both
on the softening length $\varepsilon$ and on the parameter $\eta$,
i.e. the spatial and temporal resolution are strongly entangled.  In
order to determine the role of $\eta$ we consider a set of simulations
in which all physical ($N, R_0, b_0$) and numerical parameters
($\varepsilon$, $\theta$, $\phi$) are fixed and the time-stepping
accuracy $\eta$ is varied by a factor $10^2$ (see Tab.\ref{table} for
details).

\begin{figure}
{ \par\centering \resizebox*{9cm}{8cm}{\includegraphics*{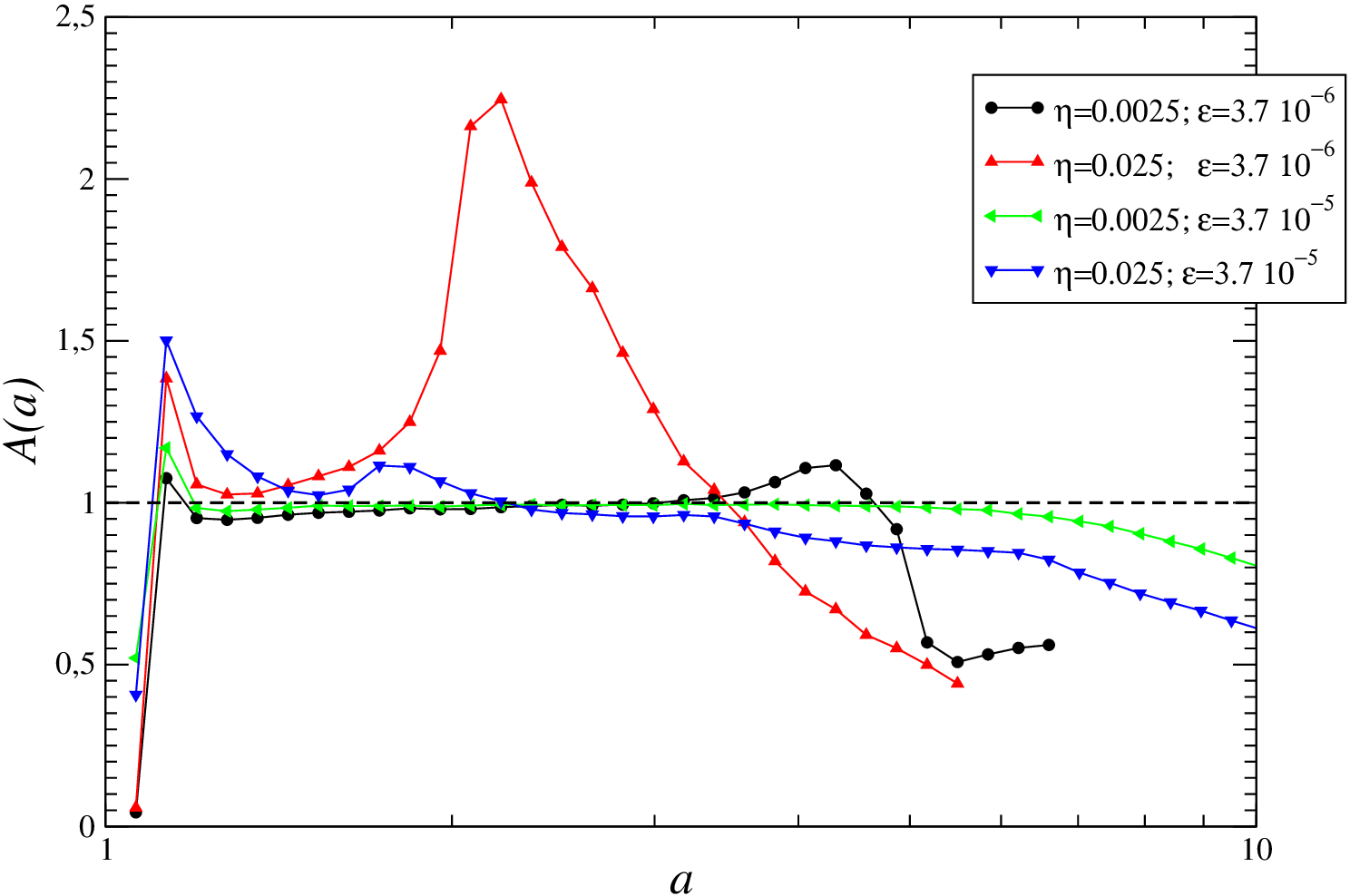}}
  \par\centering }
\caption{ Behaviour of $A(a)$ (Eq.\ref{Alpha}) for simulations with
  different $\eta$ and $\varepsilon$ (see labels) }
\label{Energy_eta} 
\end{figure}

From the LI test (see Fig.\ref{Energy_eta}) we find $a_{LI}=
a_{LI} (\eta)$.  Note that, as previously found, the LI test is
verified in about the same range of scale factors where the potential
energy is roughly constant and the structure is in the stable
clustering regime.  

\begin{figure}
\vspace{1cm} { \par\centering
  \resizebox*{9cm}{4cm}{\includegraphics*{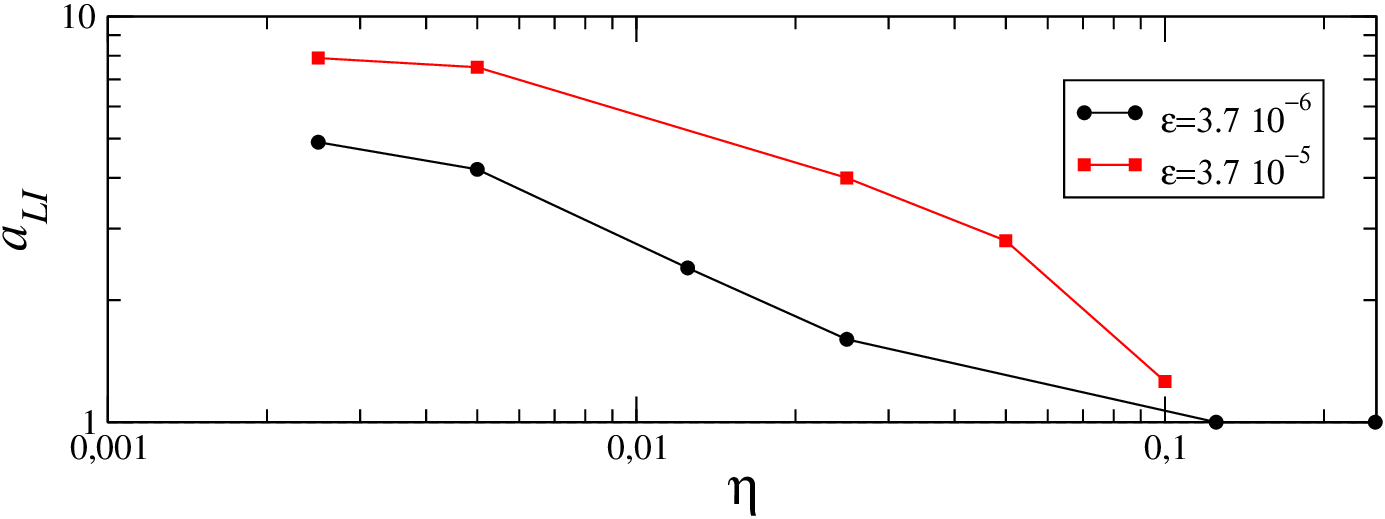}} \par\centering
}
\caption{Behaviour of $a_{LI}$ as a function of $\eta$ for different
  values of the softening length $\varepsilon$.}
\label{abreak_eta}
\end{figure}
\begin{figure}
{ \par\centering \resizebox*{9cm}{8cm}{\includegraphics*{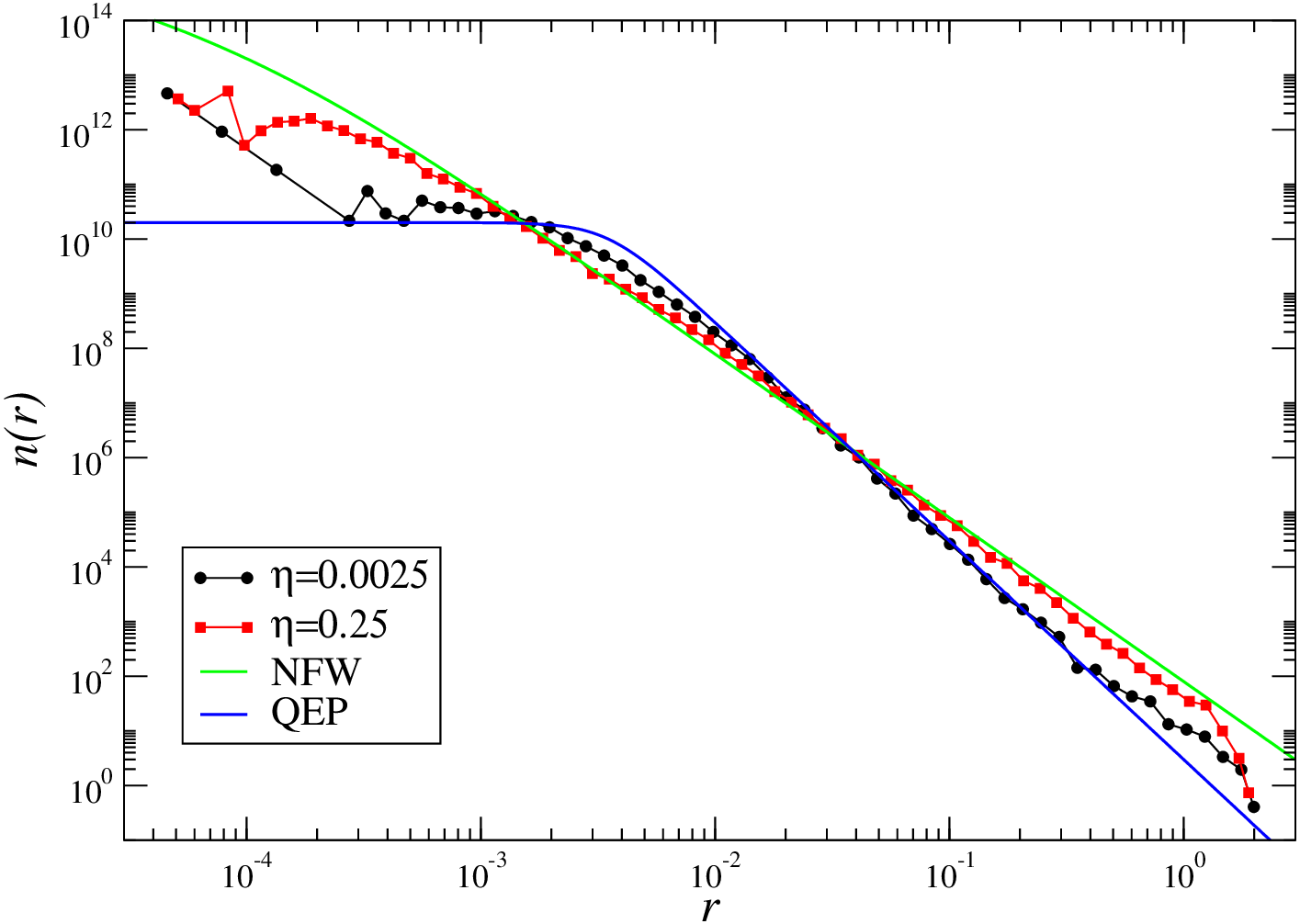}}
  \par\centering }
\caption{
Density profile in
  physical coordinates at $a=2.5$ respectively for $\eta=0.0025$  and
  $\eta=0.25$. In both the cases the softening length is
 $\varepsilon=3.7 \times 10^{-5}$.  }
\label{dp_eta_series_1}
\end{figure}

 
The behaviour of $a_{LI}$ for different values of $\varepsilon$ and
$\eta$ is reported in Fig.\ref{abreak_eta}: the larger is $\eta$ and
the sooner occurs (i) the breakdown of the LI test, (ii) the departure
from the stable clustering regime, (iii) the deviation from the QSS
with the profile given by Eq.\ref{nrb0}. That is, at fixed
$\varepsilon$, the larger is $\eta$ the smaller is $a_{LI}$. In
agreement with the results discussed above, we find that the NFW
profile provides a very good fit of the density profile for $a>a_{LI}$
(see Fig.\ref{dp_eta_series_1}).

{  The behaviours of the velocity $f(v)$ and of the nearest
  neighbours $\omega(x)$ distributions show that the smaller is $\eta$
  and the better is the resolution of close neighbours and of high
  velocity particles. In addition, the high velocity tail of $f(v)$ is
  substantially reduced in the regime $a>a_{LI}$, in agreement with
  the results obtained by changing the softening length.}


\subsection{Other numerical parameters} 

{  Results obtained by varying the opening angle $\theta$ of the
  tree algorithm from 0.7 to 0.1 and the accuracy of the relative
  cell-opening criterion $\phi$, i.e. the parameter denominated {\tt
    ErrTolForceAcc}, from $5 \times 10^{-3}$ to $5 \times 10^{-4}$ do
  not show sensible differences.  This fact gives us reasonable
  confidence that the results we have presented are quite well
  converged with respect to these parameters.  }


\subsection{Summary} 

The two main parameters that control the spatial and temporal
resolution of the numerical integration, namely the softening length
$\varepsilon$ and the time-stepping accuracy $\eta$, determine the
range of time in which a simulation is well performing, i.e., the
structure is a QSS and $W_p \approx $ const., the evolution is in
agreement with stable clustering and the LI test is reasonably well
satisfied.  Thus the scale factor at which the breakdown of the LI
test occurs is such that $a_{LI} = a_{LI}(\varepsilon,\eta) \;.$
We note that most of the discussion in the literature has focused the
attention on the role of the softening length (see e.g.
\citet{splinter_1988,heitmann_etal_2005,romeo08,joyce_2008,power_2002}):
here we have found that {\it a crucial parameter} is also represented
by the time-stepping accuracy.

Coherently with these results we found that the velocity PDF $f(v)$
and nearest neighbour distribution $\omega(x)$ depends on both the two
parameters $\varepsilon$ and $\eta$.  In particular the high velocity
tail of $f(v)$ remains stable for $a<a_{LI}$ while there is a sensible
deficit of high velocity particles for larger scale factors, due to
the limited temporal resolution. Similarly $\omega(x)$ does not change
anymore for $a>a_{LI}$, implying the departure from the stable
clustering regime. {  In this situation} the peak of $\omega(x)$ is
determined by the values of both parameters $\varepsilon, \eta$.

In summary, tests {  performed by} varying the softening parameter
show that at fixed $\varepsilon<\ell$ the smaller is $\eta$ and the
larger is $a_{LI}$. Our interpretation of this behaviour is that at a
given level of numerical accuracy, determined by the choice of
$\varepsilon,\eta$, hard scattering between particles can be resolved
only up to a certain scale factor.  Finally we note that the results
of the evolution of the density profile are coherent with the ones
found in Sect.\ref{comparison_short_long}, namely for $a < a_{LI} $
this is well approximated by the QEP shape, while for $a > a_{LI} $
the profile shows the NFW behaviour.

We have discussed that there is an artificial injection of energy into
the system during time evolution. We now conjecture that this is due
to an inaccurate integration of two-body scatterings. When an hard
scattering occurs the acceleration can be huge for a very short time
interval $\delta \tau$: however if the time stepping is larger than
$\delta \tau$ then the particle velocity will be overestimated by a
large factor: this is a known numerical problem {  discussed in
  detail by, e.g.,} \cite{knebe1999}.  Indeed, when $\eta$ decreases
the amplitude of the time stepping (see Eq.\ref{timestepacc}) gets
shorter allowing a better sampling of the change of velocity during
hard scatterings.

Instead, the change of particle velocity is $\Delta v = \Delta t
\times \mathcal{A}_{acc}$ so that for hard scatterings
$\mathcal{A}_{acc} \propto \varepsilon^{-2}$ we have $\Delta v \propto
\sqrt{\eta/ \varepsilon}$: the over-estimation of the velocity change
increases when $\varepsilon$ gets smaller. Therefore we may conclude
that the measured dependence of $a_{LI}$ on $\varepsilon, \eta$ and
$N$ is coherent with the interpretation that there is a numerical
problem in the integration of hard scatterings.
Collisionality for the smaller softenings we use manifests itself most
clearly in the breakdown of the LI test that represents a useful tool
to study these effects. 

It is interesting to note that collisional effects in cosmological
simulations have been documented in previous studies by considering
the comparison of different codes. In particular, \citet{knebe1999}
very explicitly demonstrates the effects of poor integration of hard
two body collisions at small softenings. They showed that two-body
scattering have effects which can be greatly amplified by inaccuracies
of time integration and that the effects of such poorly integrated
hard collisions is to lead to an artificial injection of energy into
the system which causes an increasing its size. {  These results are
  indeed fully compatible with our results and interpretation} (see
also the discussion in \citet{jsl13}).



\section{Role of the physical parameters} 
\label{scaling_and_finite_size} 

In this section we first study the evolution of systems with same
physical and numerical parameters but different particles number.
That is, we discretize the system with a different number $N$ of
particles taking fixed the total mass $M = m \times N$, with $m
\propto 1/N$.  Then we consider systems of different size $R_0$ but
with the number of particles, their mass and the numerical parameters
$[\varepsilon, \eta, \theta, \phi]$ fixed: this corresponds to change
only the value of $\delta$ in Eq.\ref{ratio-densities}.  Finally we
change only the initial velocity dispersion and thus the initial
virial ratio $b_0$.


\subsection{Scaling with particles number}
\label{scaling_with_N} 
 
When the number of particles is small enough, i.e. $N<N^* \approx
10^4$ we observe that during the collapse phase the LI test always
fails as $A(a)$ presents relatively large deviations from unity since
$a \gtapprox 1$ which persist at all times (see
Fig.\ref{Energy_Scaling_N_T2}).  This can be interpreted as due to the
fact that a few hard collisions occurring in the collapse phase
substantially perturb the integration of the equation of motions.
Note that according to the results obtained in
Sect.\ref{numerical_effects} the value of $N^*$ should in general
depend on $\varepsilon,\eta$.

In simulations with $N > N^*\approx 10^4 $ we find instead that
$A(a)\approx 1$ soon after the collapse phase, {with smaller
  fluctuations for larger $N$ values.  The scale factor signing the
  break of the LI test { weakly} depends on $N$,
  i.e. $a_{LI}=a_{LI}(N)$: {indeed it is found to be almost
    independent on $N$}.  This latter result seems surprising as from
  Eq.\ref{tau2} the collisional relaxation time should {increase} with
  $N$ so that one would naively expect that larger is $N$ and larger
  is $a_{LI}$.

To interpret this behaviour we recall that the density of the core
after the collapse, $n_c$ in Eq.\ref{nrb0}, scales with the number of
particles as $n_c \propto N^2$: at fixed mass density, by increasing
the number of particles, the density of the core increases as $\rho_c
= n_c \times m \propto N$, given that particles mass scales as $m
\propto N^{-1}$ \citep{jmsl09a}.  From Eq.\ref{tau2} we thus find that
$\tau_2 = \kappa N \tau_c \propto \kappa \sqrt{N}$, as $\tau_c\approx
1/\sqrt{G\rho_c}$. Given that $\kappa$ is weakly dependent on $N$,
i.e. $\kappa \propto 1/\log(N)$, we thus find that $\tau_2$ is almost
independent on $N$, at least for the limited range of $N$ we
considered, i.e. $N=10^4\div 10^5$.  
\begin{figure}
\vspace{1cm} { \par\centering
  \resizebox*{9cm}{8cm}{\includegraphics*{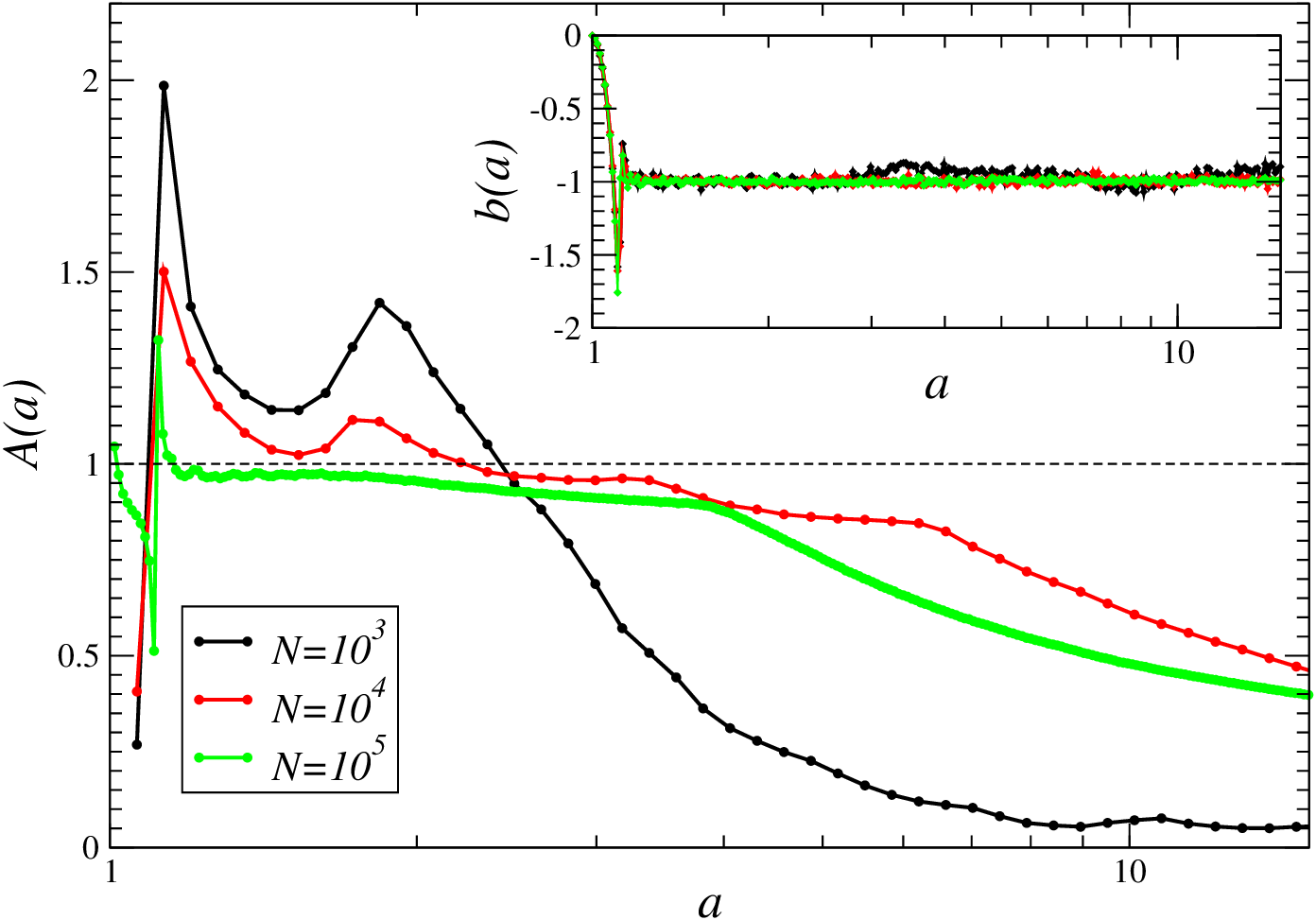}} \par\centering
}
\caption{Behaviour of $A(a)$ (Eq.\ref{Alpha}) for
  simulations with different $N$ (see Tab.\ref{table}) Insert
  panel: Behaviour of the virial ratio for the same simulations} 
\label{Energy_Scaling_N_T2} 
\end{figure}

In Fig.\ref{dp_N_series_1} it is shown the density profile in physical
coordinates for simulations with $N=10^3, 10^5$ particles: note that
$n(r)$ has been normalised by recalling the results of \citet{jmsl09a}
where it was shown that the two parameters in Eq.\ref{nrb0} scale as
$r_c \approx N^{-1/3}$ and $n_c \approx N^2$.  As we found in the
previous sections, the breakdown of the stable clustering regime
corresponds to a deformation of the profile.
\begin{figure}
\vspace{1cm}
{
\par\centering \resizebox*{9cm}{8cm}{\includegraphics*{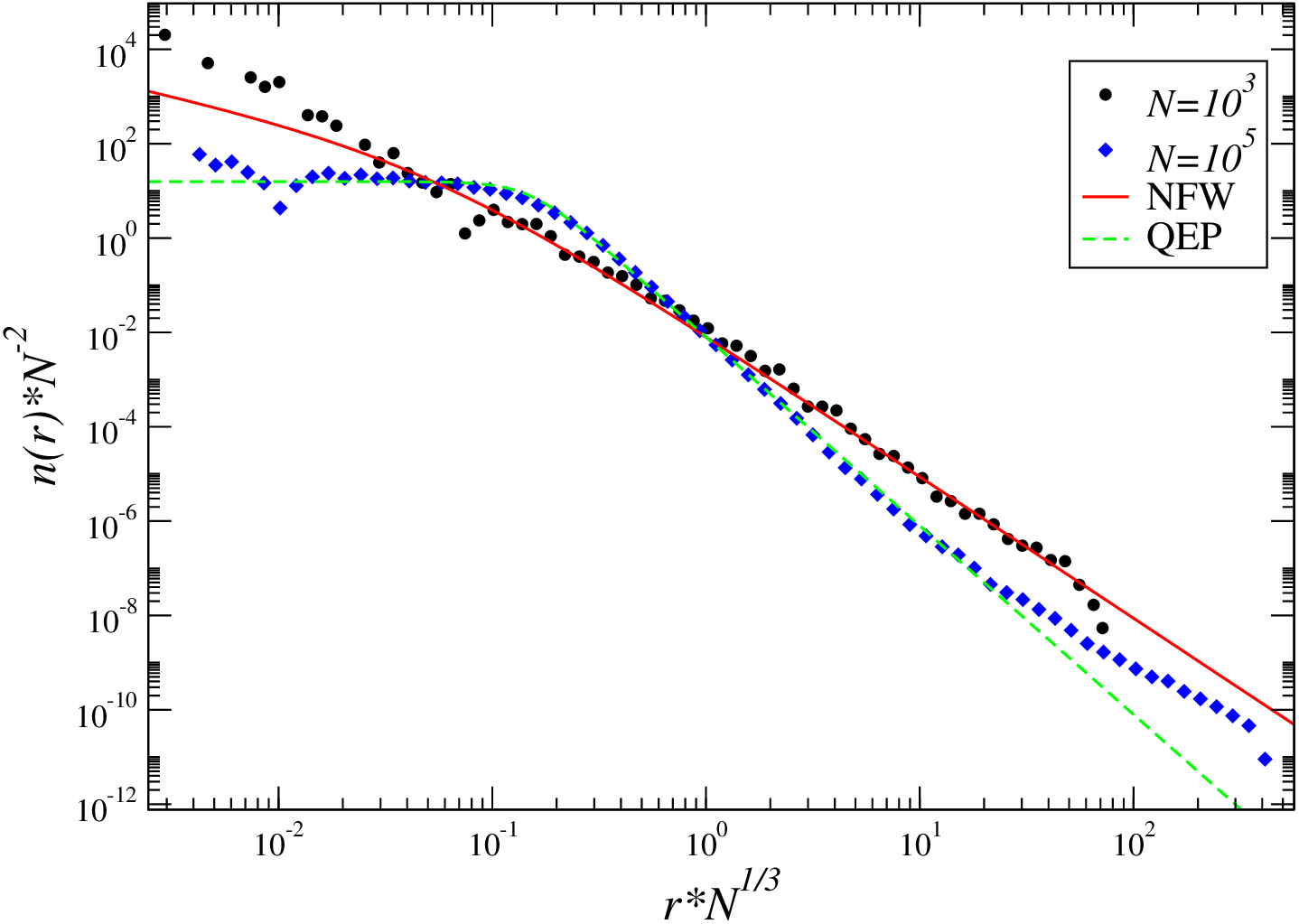}}
\par\centering
}
\caption{Density profile in physical coordinates for simulations with
  $10^3$ and $10^6$ particles, at $a=4.9$, renormalised as explained
  in the text at different values of the expansion factor.  The best
  fit with Eq.\ref{nfw_eq} (NFW) and Eq.\ref{nrb0} (QEP), respectively
  for $N=10^3$ and $N=10^5$ are reported as reference.}
\label{dp_N_series_1}
\end{figure}
{  In the case shown in Fig.\ref{dp_N_series_1}, while for $N=10^5$
  the simulation is still in the QSS regime, for the $N=10^3$ case the
  density profile is already well fitted by the NFW shape (see
  Fig.\ref{dp_N_series_1}).}


\subsection{Finite size effects}
\label{finite_size_effects}

By considering the behaviours of the gravitational potential energy
and of the LI test for simulations with different initial size $R_0$
we can conclude that $a_{LI}= a_{LI}(R_0).$ In particular, the smaller
is $R_0$ and the smaller is $a_{LI}$.
\begin{figure}
\vspace{1cm}
{
\par\centering \resizebox*{9cm}{8cm}{\includegraphics*{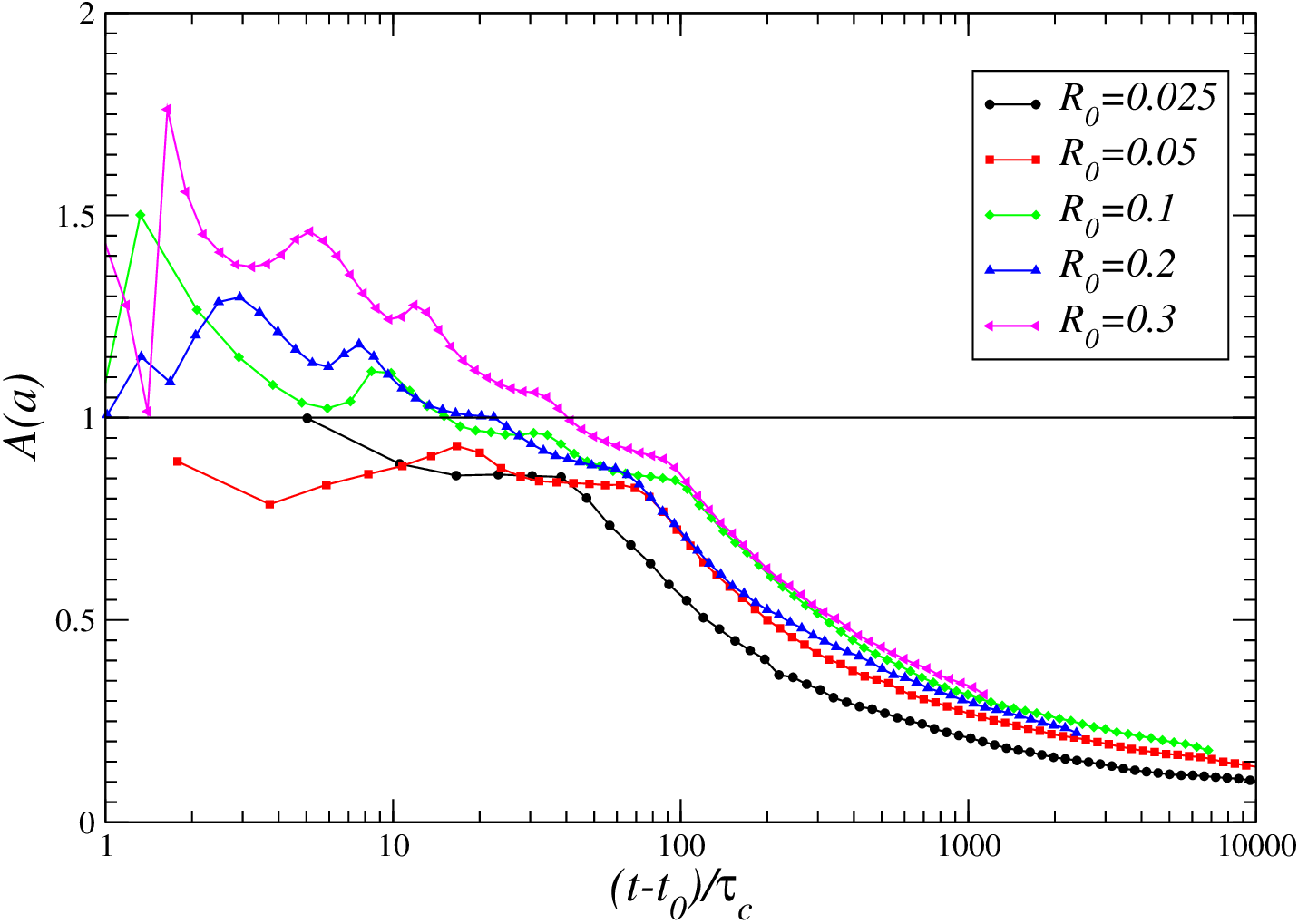}}
\par\centering
}
\caption {Behaviour of $A(a)$ (Eq.\ref{Alpha}) for simulations with
  different initial size $R_0$ as a function of time (normalised to
  $\tau_c(R_0)$).  }
\label{Energy_Radius_T2_b} 
\end{figure} 
This fact can be simply interpreted by considering that when the
system size decreases, the amplitude of the over-density
$\delta=\delta(R_0)$ increases and so the value of the characteristic
time scale $\tau_c=\tau_c(R_0)$ (see Tab.\ref{tableErrE}). Thus, to compare the
behaviours for different $R_0$ we plot (see
Fig.\ref{Energy_Radius_T2_b} --- bottom panels) $W_p(a)$ and $A(a)$ as
a function of time in units, for each value of $R_0$, of the
corresponding value of $\tau_c=\tau_c(R_0)$.  The residual difference
in the observed behaviours is due to the effect of the different ratio
$R_0/L$: when $R_0 > 1/10 L$ finite size corrections, {i.e. the
  terms neglected in approximating Eq.\ref{3d-equations-1} with
  Eq.\ref{3d-equations-4} become important} \citep{jsl13}.

The evolution of the density profile in physical coordinates for the
systems with different $R_0$ (see Fig.\ref{dp_R0_T2_series_1}) shows
that they firstly reach the QEP behaviour; then the density profile
approaches the $r^{-3}$ behaviour, at different values of the scale
factor for different $R_0$ and in a coherent way with the behaviour of
$A(a)$.
\begin{figure}
\vspace{1cm}
{
\par\centering \resizebox*{9cm}{8cm}{\includegraphics*{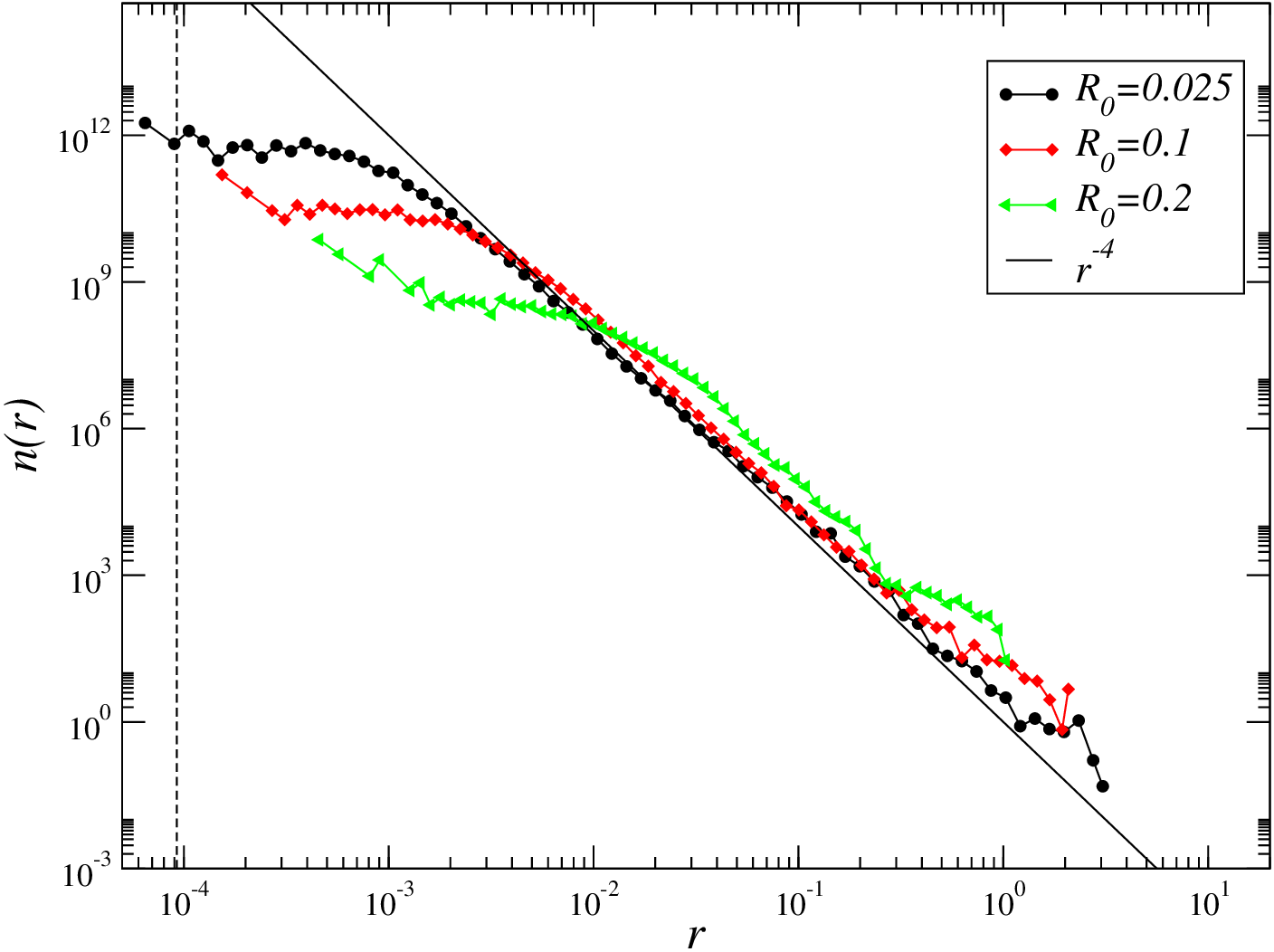}}
\par\centering
}
\caption{Density profile in physical coordinates at $a=2.5$ for
  different values of $R_0$ (see labels). The vertical dashed line
  correspond to $a\times \varepsilon$. A reference line with slope
  $-4$ is plotted as reference.}
\label{dp_R0_T2_series_1}
\end{figure}

Finally we note that for long enough values of the scale factor 
the density profile for different
$R_0$ converges precisely to the same behaviour characterised by a
$x^{-3}$ decay.  As discussed in Sect.\ref{numerical_effects}, at
fixed time-stepping accuracy, the small scale properties of the
distributions are determined by the value of the softening length.
Indeed, independently on the initial value of $R_0$, for long enough
times, all curves collapse on each other.
\begin{figure}
\vspace{1cm}
{
\par\centering \resizebox*{9cm}{8cm}{\includegraphics*{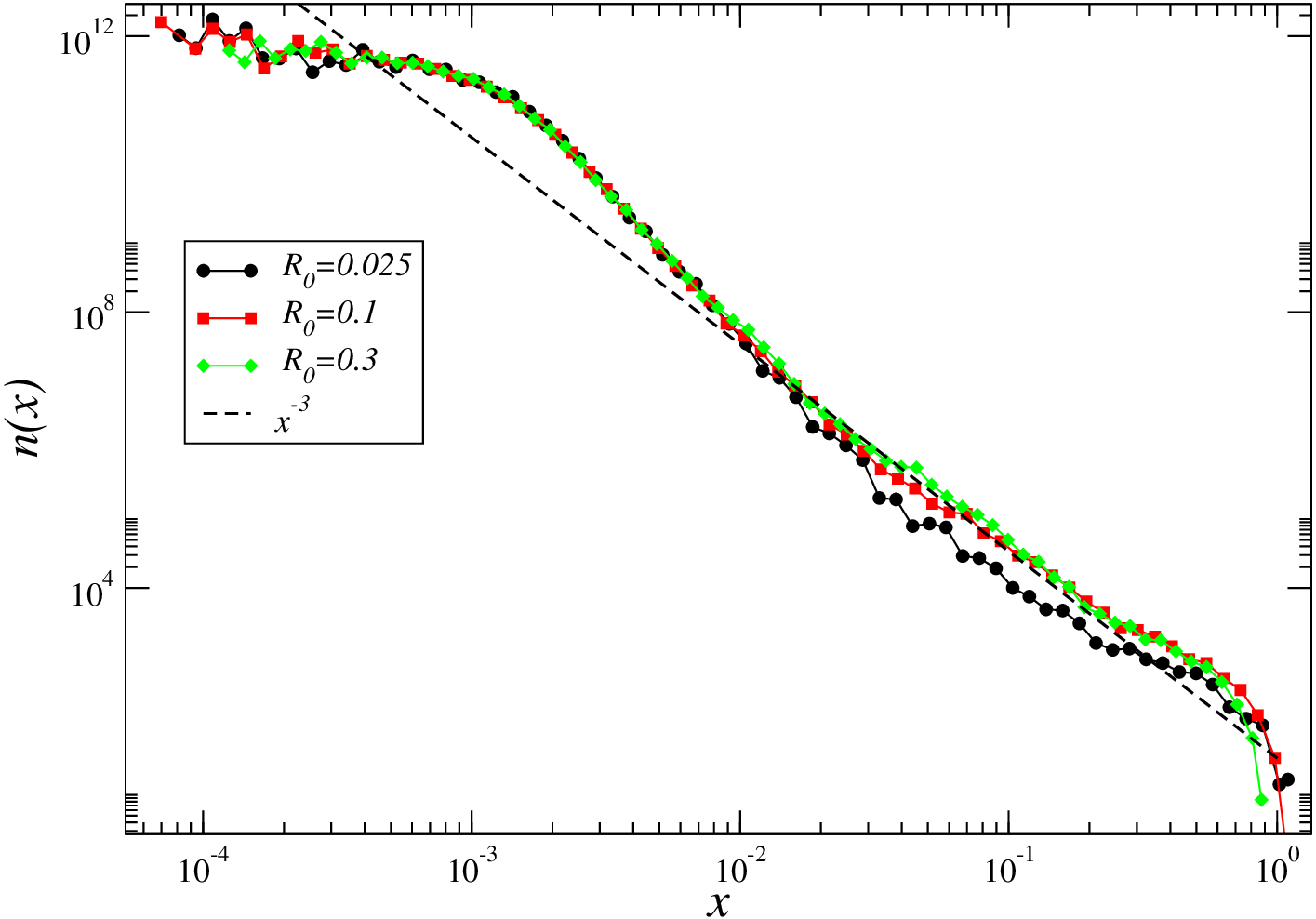}}
\par\centering
}
\caption{Density profile in comoving coordinates at $a=70$ for the
  simulations with $\varepsilon = 3.7 \times 10^{-4}$ and a different
  value of the initial radius $R_0$ (see labels).  A reference line
  with slope $-4$ is plotted as reference.}
\label{dp_T2T3_series_radius}
\end{figure} 


\subsection{Role of the initial velocity dispersion}
\label{initial_velocity_dispersion}

The initial conditions we have considered can be thought to represent
an halo which virializes in a relatively short time and which then
evolve in isolation from the rest of the mass of the universe. Indeed,
when $b_0 \rightarrow -1$ the halo is already almost virialized while
when $b_0 \rightarrow 0$ the halo is far from such a
configuration. However for $b_0\in [-1,0]$ the structure undergoes to
a collapse, which can be less or more violent, that drives the system,
in about the same time-scale $\tau_c$ (see Eq.\ref{tauC}), into a
virialized quasi-stationary state that can be well described as a
collision-less equilibrium configuration \citep{sl12a}.

Fig.\ref{Energy_Virial_T2} shows the behaviour of $A(a)$ for
simulations with different initial virial ratio $b_0 \in [-1,0]$,
i.e. with different initial velocity dispersion.  We find that
$a_{LI}=a_{LI}(b_0)$: in particular, the smaller is $b_0$ the larger
is $a_{LI}$. This behaviour is easily explained by considering that
the closer the system is initially to the virial configuration and the
less violent is the dynamics that drives the system into the
virialized QSS \citep{sl12a}: the larger is $b_0$ at the higher is the
density of the core reached after the collapse.
\begin{figure}
\vspace{1cm}
{
\par\centering \resizebox*{9cm}{8cm}{\includegraphics*{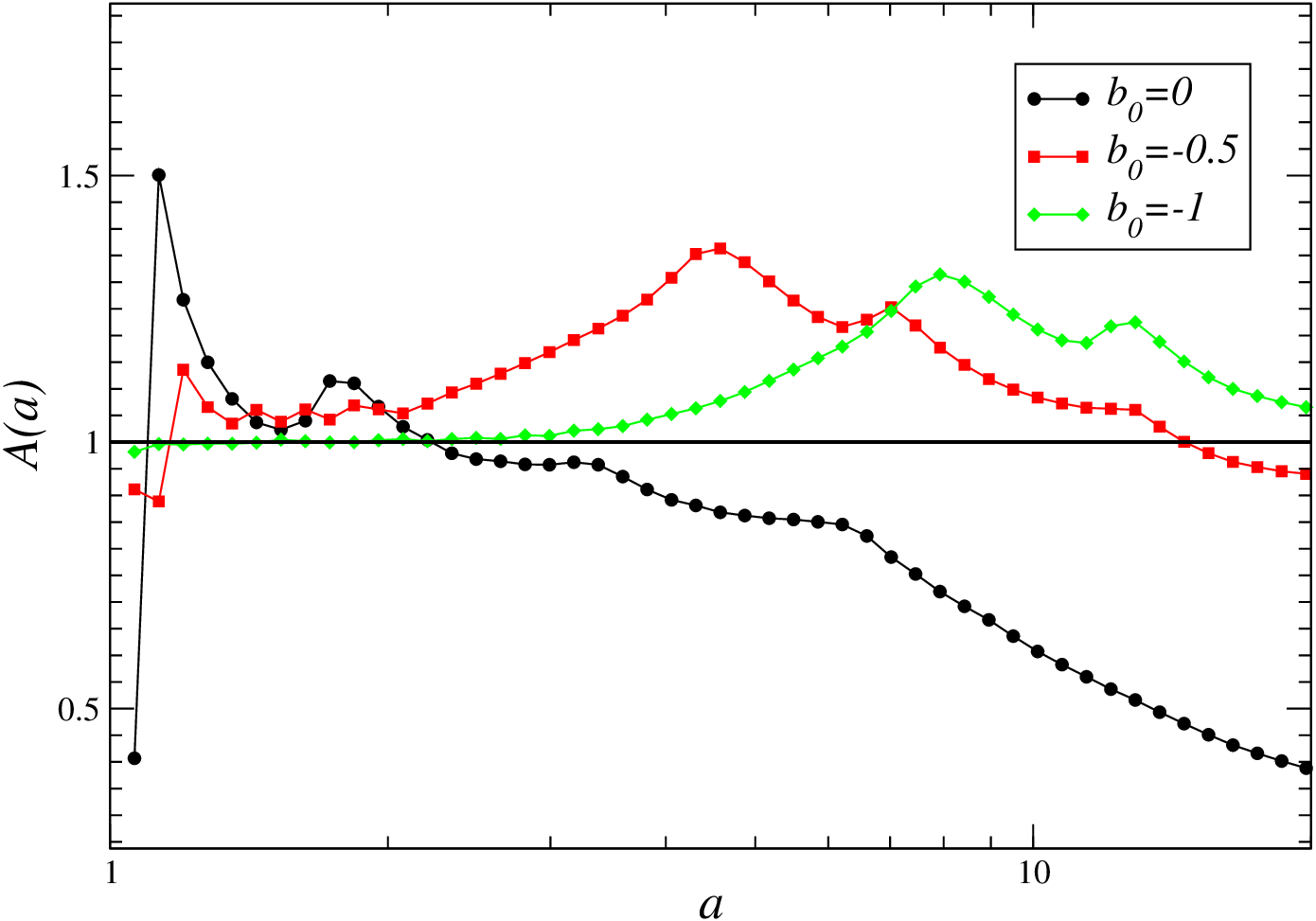}}
\par\centering
}
\caption{ Behaviour of $A(a)$ (Eq.\ref{Alpha}) for for simulations
  with different initial size $b_0$.  }
\label{Energy_Virial_T2} 
\end{figure} 

We recall that, in the STAT case, the formation of the $r^{-4}$ is
observed only when the collapse is violent enough that a fraction of
the initial mass and energy is ejected from the system at the
formation of the QSS: in \citet{sl12a} it was found that this occurs
for $b_0 > -1/2$.  In this situation the system evolution is very
similar to the $b_0=0$ case: at the beginning the density profile is
well approximated by the QEP and then, for $a>a_{LI}$ it develops a
$r^{-3}$ tail approaching the NFW shape also at small radii.  Thus,
although the $b_0=0$ is peculiar because the collapse is instantaneous
in the continuous limit, {\it it does not represent a pathological
  case} and the $r^{-4}$ tail is formed as long as the collapse is
violent enough that part of the mass and energy is ejected from the
virialized structure.

For values of the initial virial ratio $b_0<-1/2$ the virialized
structure after the collapse does not present a density profile with a
$1/r^4$ tail. { However for long enough times the density profile
  approaches the $1/r^3$ behaviour for any value of $b_0$. This shows
  that the $1/r^3$ is formed through the dynamical process and that it
  is thus independent on the initial conditions. }

In Fig.\ref{DP_CC_Vir_0.com} it is shown the comparison of the density
profile, in comoving coordinates, for simulations with different
initial virial ratio and at different expansion factors: this confirms
that the smaller is $b_0$ and the slower is the evolution. In
particular, the smaller $b_0$ the longer it takes to form the $1/r^3$
tail, a behaviour that is approached at long enough times for any
value of $b_0$.
\begin{figure}
\vspace{1cm} { \par\centering
  \resizebox*{9cm}{8cm}{\includegraphics*{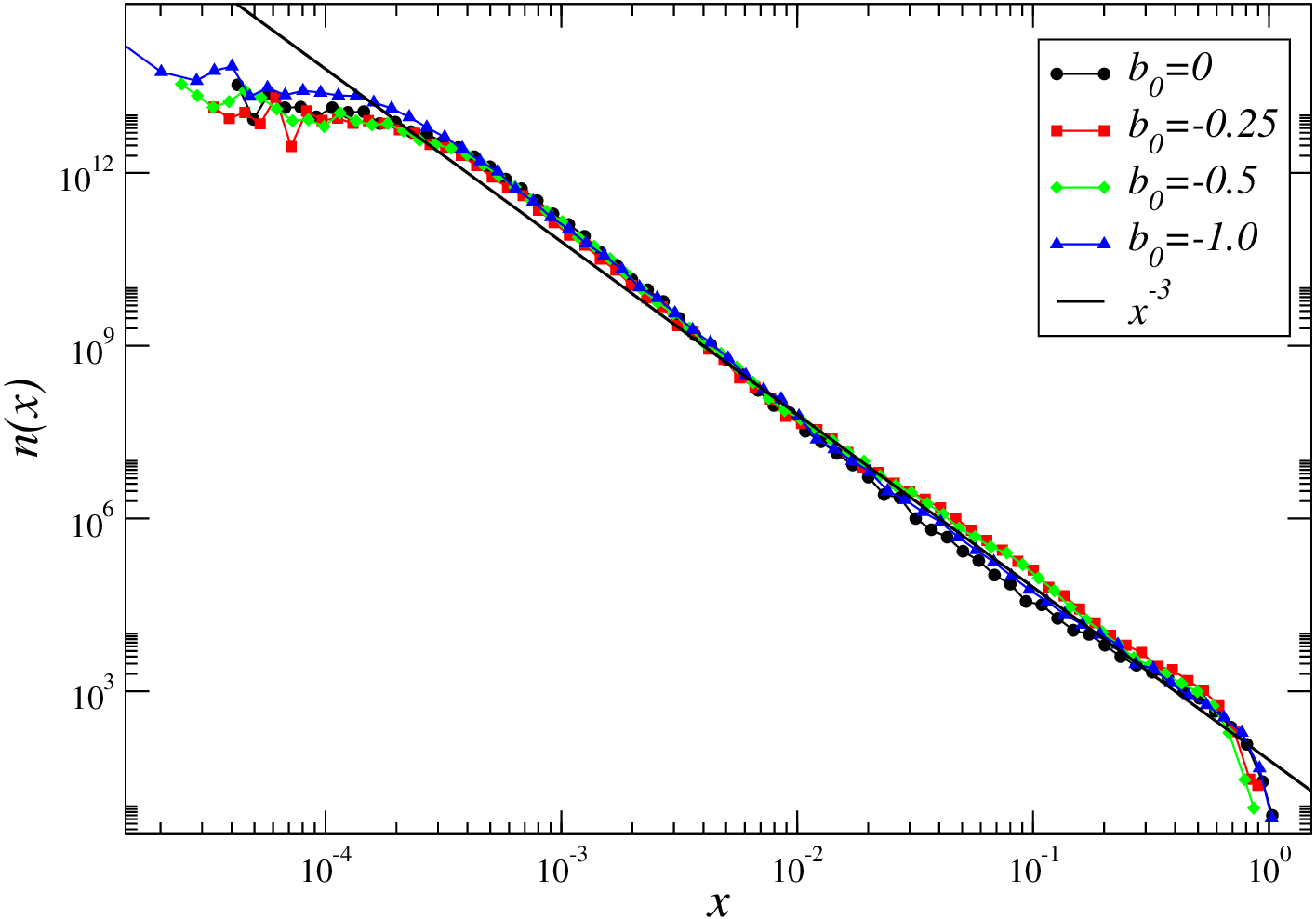}} \par\centering
}
\caption{Comparison of the density profile in comoving coordinates for
  simulation with different initial virial ratio a $a=70$.}
\label{DP_CC_Vir_0.com}
\end{figure}


\subsection{Summary} 

In summary we found that $a_{LI} = a_{LI}(R_0,N,b_0)$, i.e.  the
regime in which the system evolution is correctly numerically
integrated depends on the physical parameters of the initial
conditions.

For what concerns the dependence of $a_{LI} $ on $N$, we have noticed
that: (i) for relatively small particles number, i.e. $N<N^*\approx
10^4$, the system never reaches the quasi-stationary state in the
stable clustering regime. We have interpreted this behaviour as due
the fact that a few hard scatterings are enough to perturb the
macroscopic dynamics.  {(ii) On the other hand for large enough
  particles number, i.e., $N \in [10^4,10^5]$, we find that the
  deviation from the QSS slowly depends on $N$. This occurs because
  when initial conditions are cold, the system contracts more for
  larger $N$ values, so that the virialized state has a core mass
  density $\rho_c \propto N$ and thus the typical time scale for
  crossing is $\tau_c \propto 1/\sqrt{N}$: in this situation the
  collisional time scale $\tau_2$ (see Eq.\ref{tau2}) is {  very
    weakly dependent } on $N$.}

By decreasing the system size $R_0$ we find that $a_{LI}$ decreases as
well: this is coherent with the fact that the smaller is $R_0$ and the
smaller is the average distance between nearest neighbour particles in
the core, and thus the higher the probability of hard collisions.
However when we renormalize the behaviours to the proper
characteristic time $\tau_c=\tau_c(R_0)$ we find that the differences
in the departure from the stable clustering regime become smaller:
these residual differences are due to finite size effects. 

In addition we have found that for long enough times all systems with
initially different $R_0$ converge to the same density profile
characterised by a $r^{-3}$ decay. This convergence is obtained
because the numerical integration is badly performed, as shown by the
LI test.

Finally, by increasing the initial velocity dispersion, the breakdown
scale factor $a_{LI}$ is found to increase. This behaviour can be
simply explained by considering that when $b_0 \rightarrow 0$ the
collapse is much more violent than for $b_0 \rightarrow -1$; thus { 
  in the former case} the core may reach higher densities sooner, then
having a similar evolution to the cold case. In addition, when the
initial velocity dispersion is high enough, the system remains in the
stable clustering regime longer and until the softening length, that
increases in physical coordinates, becomes large enough to perturb the
evolution. At later times the system shows an evolution that is again
similar to the cold case.


\section{Discussion and Conclusion}
\label{conclusion}

In this paper we have tested a cosmological code { ({\tt GADGET})} by
choosing very simple and idealised initial conditions. While we do not
claim that this choice can be justified by any physical argument, {
  i.e. the toy model does not represent any realistic physical
  system}, we stress that { in this way we have been able } to make a
series of controlled tests on the accuracy of a cosmological
code. Given what we have learnt from this case, we discuss here some
general ideas for the testing of a cosmological code in the case of
more realistic initial conditions, as those predicted by standard
models of galaxy formation. In order to consider separately these two
different but, as we argue below, probably related issues, we first
discuss the results of the simulations we have presented in this paper
and then we briefly comment on the more complicated problem that
requires more tests and work to be properly explored.

\subsection{The control of the accuracy of a cosmological code: a toy model}

The evolution of an isolated over-density {in a cosmological N-body
  simulation} is {  approximately} equivalent, in physical
coordinates, to the evolution of the same structure in an open system
without expansion \citep{jsl13}.  This equivalence is strictly
verified in the limit of vanishing force smoothing and when the size
of the over-density is much smaller than the size of the simulating
box.  When these conditions are satisfied, the evolution of the open
system can be used as a template to compare the evolution in an
expanding background with the aim of determining the accuracy of a
cosmological simulation.  The open system evolution is such that it
relaxes to a virialized quasi stationary state (QSS) in a time
$\tau_c$ (see Eq.\ref{tauC})
\citep{vanalbada_1982,aarseth_etal_1988,jmsl09a,sl12a} and it
evaporates, due to collisional effects, in a time $\tau_2 \gg \tau_c$
(see Eq.\ref{tau2}).  {  The QSS corresponds, in a cosmological
  N-body simulation, to an isolated structure in the stable clustering
  regime.}

\citet{jsl13} considered the evolution of an isolated over-density
initially close to virial equilibrium: for this reason during the
collapse its size and density vary a little. In this paper we
considered {  a more general case, namely} an isolated over-density
with initial virial ratio in the range $b_0- \in [-1,0]$. When $b_0
\rightarrow -1$ the initial conditions are warm enough that the
collapse consists in a relatively small modification of system
size. Instead when initial velocities are cold, i.e. $b_0 \rightarrow
0$, the system largely contracts during the collapse, reducing
considerably its size {and shape} and thus rapidly reaching, {  in
  its central core}, a very high density \citep{sl12a}.

{  Note that an} initially cold, i.e. $b_0=0$, and spherical
over-density {  does not represent} a pathological case {  from the
  dynamical point of view}: indeed, as firstly noticed by
\citet{stiavelli1987}, a density profile with a $r^{-4}$ tail is a
generic property of systems formed by a violent relaxation
process. For instance, recently \citet{sl12a,sl12b} found that this
profile is formed also in the collapse of an initially spherical and
isolated cloud of particles with a small enough velocity dispersion,
with a uniform or power-law density profiles.

We found that, for $b_0 \rightarrow 0$, once the system has reached a
virialized configuration, {its evolution can be} characterised by two
different regimes:
\begin{itemize}
\item (i) for scale factors $a<a_{LI}$ the potential
energy in physical coordinates $W_p$ is approximately constant, the
Layzer Irvine (LI) equation, describing the energy change in an
expanding background, is {  well} satisfied, the density profile is
characterised by a $r^{-4}$ tail, characteristic of the quasi
equilibrium profile (QEP) in the open case (see Eq.\ref{nrb0}), and
the system is in the stable clustering regime, which means that it
remains stable in physical coordinates. 
\item (ii) On the other hand for
$a>a_{LI}$ we find that $|W_p| \sim a^{-1}$, there is a breakdown of
the LI equation, the density profile becomes {\it stable in comoving
  coordinates}, which corresponds to the breakdown of the stable
clustering regime, it develops a $r^{-3}$ tail and it can reasonably
be fitted by the NFW profile.
\end{itemize}

In principle, the QSS has a lifetime $\tau_2$ that diverges with the
number $N$ of system particles because of collisional effects.
However we found that the time scale corresponding to $a_{LI}$ is
(i) $t_{LI}<\tau_2$ and (ii) it depends on the numerical
parameters of the simulation.  Therefore this time scale is a
numerical artifact: the relevant question is what determines it.

In order to answer to this question, we have performed various series
of simulations by varying the integration parameters, such as the
softening length of the gravitational force $\varepsilon$ (which, { 
  in the simulations we have performed,} is fixed in comoving
  coordinates during the evolution) and the accuracy of the
  time-stepping $\eta$. We found that
\[
a_{LI} = a_{LI}(\varepsilon,\eta). 
\]
In particular, we found that there is a narrow range of $\varepsilon$
(at fixed $\eta$) for which the simulation behaves ``well''
(i.e. regime (i) above). When $\varepsilon $ is much smaller than the
initial inter-particle distance $\ell$ the integration is more
difficult and $a_{LI}$ decreases: this fact can be interpreted as due
the effect of hard collisions which require a high numerical accuracy
to be correctly integrated. On the other hand, when $\varepsilon$ is
of the order of the characteristic scale of the virialized system the
mean field potential is no longer Newtonian {  so that the system
  deviates from the virialized configuration and from the quasi
    equilibrium profile.}

That there is a problem in the accuracy of the numerical integration
is confirmed by varying $\eta$ at fixed $\varepsilon$ (with
$\varepsilon < \ell$): in particular when $\eta$ decreases $a_{LI}$
increases.  This fact can be again interpreted in terms of a better
accuracy of the numerical integration and in particular a more
accurate integration of high velocity particles, which require a
smaller time-stepping to be properly followed.  Both tests by varying
$\varepsilon$ and $\eta$ are {  thus} consistent with the interpretation
outlined above, that confirms the results by \cite{jsl13}: two-body
scatterings perturb the accuracy of the simulation evolution.  Thus
the temporal and spatial resolution of the simulation are entangled
and both must be considered for the determination of the optimal
values of the numerical parameters.

{  Note that codes with adaptive spatial resolution (see e.g.,
  \citet{knebe1999,teyssier,miniati}) may perform much better than the
  fixed resolution case which we have considered in this
  paper. Indeed, it is clear such codes will, by construction, avoid
  respect. The tests we have discussed in this paper can be easily
  applied to other codes and a detailed study of codes with an
  adaptive mesh refinement is postponed to a forthcoming work.}

{  Because of the violation of the LI, for $a>a_{LI}$, the system
  expands in physical coordinates, thus increasing its potential
  energy: this is due to an injection of energy into the system}. 
From the tests performed by varying the softening length and the
time-stepping parameter we conclude it is the effects of poor
integrated hard two body collisions that leads to an artificial
injection of energy into the system.  This same conclusion was drawn
by \citet{knebe1999} by considering a comparison of different N-body
codes.  In summary we find, in agreement with \citet{jsl13}, that
given a certain choice of numerical parameters, one may follow
structure formation in the non linear regime, for a limited time which
depends on several physical properties of the system. The results
obtained here clearly show the role of the numerical (in)accuracy in
the determination of the shape of the density profile structure.

In order to further test the result that the numerical integration
becomes inaccurate because of collisional effects, we have, at fixed
numerical parameters, changed the physical parameters of the initial
conditions. Namely we have varied the number of particles $N$ (at
fixed density), the size of the structure $R_0$ and the initial
velocity dispersion (and thus the initial virial ratio $b_0$). As
result we found that
\[ a_{LI} = a_{LI}(N,R_0,b_0) \;. \]
When decreasing $R_0$ we find that $a_{LI}$ decreases: the initial
inter-particle distance gets smaller, as $\ell \propto R_0$, and thus
hard scatterings are more probable.  The dependence of $a_{LI} $ on
the particle number clearly shows the effect of two-body
collisions. For small particle number, i.e. $N<10^4$, the system never
reaches a QSS and the LI test fails at all times. We then find that,
at fixed $\varepsilon$ and $\eta$, {and for $N\in [10^4, 10^5]$ the
  breakdown expansion factor $a_{LI}$ is almost independent with $N$.
  This behaviour can be interpreted as due to the fact that the larger
  is $N$, the larger the core mass density and thus the characteristic
  time $\tau_c$: in this situation the collisional time scale $\tau_2$
  in Eq.\ref{tau2} becomes almost independent on $N$.}

Finally by increasing the initial velocity dispersion, i.e.
decreasing $b_0$, we find that $a_{LI}$ increases: indeed, as shown
in \citet{sl12a}, when the initial configuration is closer to the
virial equilibrium, i.e. $b_0=-1$, the collapse is less violent so
that the size and density of the structure after the collapse are not
largely changed as it occurs when $b_0=0$.  In this latter case the
core of the structure after the collapse may reach very high
densities, i.e.  a more favourable condition for the occurrence of
two-body scatterings. However, we have found that for $b_0 \in [-1,0]$
the density profile, in the long time evolution, approaches the same
NFW shape. { Thus, the density profile, for long enough times,
  approaches the $1/r^3$ behaviour for any value of $b_0$ a fact that
  shows that this tail is formed by the same dynamics process,
  independent on the initial conditions, in all cases. }

While the origin of the $r^{-4}$ can be understood by considering that
the energy of particles orbiting around the core is conserved
\cite{sl12a}, the effect of inaccurately integrated two-body
scatterings is indeed that of perturbing such energy conservation.  {
  The energy injection, is especially efficient for the fastest
  particles, i.e. those that form the tail of the density profile. A
  quantitative study of this effect is postponed to a forthcoming
  work, but qualitatively we may conclude that the deformation of the
  density profile is related to the inability to resolve numerically
  high velocity particles and thus to conserve their energy properly.}


\subsection{Some comments on more complex initial conditions}

{  The simplicity of the system considered as initial conditions has
  let us to develop systematic tests to control the accuracy of a
  cosmological code.  The principal result that we have obtained is
  that the range of redshifts where the numerical integration of a
  certain structure is reliable, i.e. it follows the stable clustering
  regime and it passes the LI test, is in general limited to narrow
  one that is dependent on the various physical (number of particles,
  velocity dispersion, density and amplitude of the over-density) and
  numerical (softening length, time-stepping) parameters of a given
  simulation.  Therefore, our results suggest that the limited
  numerical resolution of a given simulation can be not sufficient to
  integrate the formation of structures with a wide range in size and
  mass as those formed in the currently favoured cosmological models
  structures through a hierarchical aggregation. A more detailed study
  of the more complex cosmological case is thus required.}

{  Here we note that}, if one is not able to integrate correctly the
evolution of this simplified system, one will encounter more difficult
problems if one wants to properly simulate the formation of structures
of different sizes, with different particle numbers and in a complex
environment as that occurring in the case of typical initial
conditions of structure formation models.  {  On the other hand} the
choice of the physical and numerical parameters adopted in this paper
could be completely unrealistic and thus irrelevant for the case of a
cosmological simulations that uses initial conditions as those
predicted by standard models of galaxy formation.

{  In this respect, we note that the numerical parameters have been
  varied in a relatively large range, i.e. $\varepsilon \in [10^{-6},
    10^{-3}]$ and $\eta \in 2.5\times [10^{-3},10^{-1}]$.  In
  addition, the number of particles used, ranging in $N \in
  [10^3,10^5]$, is comparable to the number of particles which form a
  single halo in a typical cosmological simulation \citep{millennium}.
  Thus the system we considered, once it reaches a virialized state,
  can be though to represent a simplified toy model approximating a
  virialized cosmological halo at its formation time. This structure,
  however then evolves as if it were isolated from the rest of the
  universe.  Thus, such a self-gravitating system presents several
  differences with respect to the halos formed in typical cosmological
  simulations: it is spherical and isolated so that it has initially a
  simple geometry, it is not subjected neither to mass accretion nor
  to tidal interactions with the neighbouring structures, it does not
  fragment into (macroscopic) substructures during its evolution and
  it does not collide with structures of comparable size.}

A possible method to control both the role of the physical and the
numerical parameters {  in a more complex cosmological simulations}
was presented in \citet{jsl13}, where it was outlined a sort of recipe
that can be simply implemented and that can furnish a general manner
for testing the accuracy of a simulation of halos containing $N$
particles. This is simply based on the {  testing} of whether an
halo formed in a complex environment, once it is isolated, evolves in
agreement with stable clustering and represents a quasi-stationary
configuration, given the same numerical parameter of the numerical
integration (softening length, time-stepping, etc.)  and the same
physical parameters of the structure (number of particles, size and
velocity dispersion). {  The implementation of this test is postponed
  to a forthcoming work.}

As a final remark, we stress that we have observed that the density
profile of the virialized structure formed from the simple initial
conditions we have considered converges, for a wide range the
numerical and physical parameters of the simulation, to the NFW
profile, i.e. the same profile observed in cosmological N-body
simulations that start from the typical initial conditions predicted
by standard models of galaxy formation. Whether this is a coincidence
or whether also standard cosmological halos are seriously affected by
the numerical (in)accuracy of the numerical integration is an open
problem that will explored in full details in forthcoming works

\bigskip

I thank Michael Joyce for useful discussions and comments, Roberto
Ammendola and Nazario Tantalo for the valuable assistance in the use
of the Fermi supercomputer where the simulations have been performed.
{  I also would like to thank Alessandro Romeo for a number of very
useful comments and suggestions that have allowed to improve the
presentation of the paper.}


\bibliographystyle{mn2e}

\end{document}